\documentclass[notitlepage,12pt]{report}
\usepackage[utf8]{inputenc}
\usepackage{amsmath}
\usepackage{commath}
\usepackage{latexsym}
\usepackage{amssymb}
\usepackage{bm}
\usepackage{spalign}
\usepackage{wrapfig}
\usepackage[utf8]{inputenc}
\usepackage{fancybox}
\usepackage{natbib}
\usepackage{bbold}
\usepackage{graphicx}
\usepackage{float}
\usepackage{caption}
\usepackage{subfigure}
\usepackage{color}  
\usepackage{titling}
\usepackage{blindtext}
\usepackage{hyperref}
\hypersetup{colorlinks=true,linkcolor=blue,citecolor=blue}
\graphicspath{ {images/} }
\hypersetup{linktocpage}

	\newsavebox{\abstractbox}
	\renewenvironment{abstract}
	{%
		\global\setbox\abstractbox=\vtop\bgroup
		\begin{center}\bfseries\abstractname\end{center}%
	}
	{\par\egroup}

	\title{\textbf{Josephson junctions with spin-orbit
			and spin-flip interactions}}
	\author{\textbf{A. F. Tzortzakakis}, \textbf{N. Flytzanis}\\	Physics Department, University of Crete, Heraklion, 71003, Greece}
	\date{}
	
\begin{document}
		\setcitestyle{numbers}
		\begin{titlingpage}
			\begin{abstract}
				In this thesis we study short Josephson junctions which include a region with Rashba spin-orbit coupling effect. Our junctions consists of two superconductors (S) and a 2-dimensional electron gas (2DEG) layer between them: S/2DEG/S junction. We also include two thin insulating interfaces between the superconductors and the 2DEG, which are capable of both normal and spin-flip scattering. The junctions we study are assumed to be in the ballistic limit and so, we do not consider the effects of impurities. The basic equations we use for our model are the Bogoliubov-de Gennes equations.
				\par We give particular emphasis in the relation between the supercurrent of the junction as a function of difference of the phase parameters of the two superconductors. We study thoroughly how this relation differs to the change of the junction's length, the spin-orbit coupling constant, the normal scattering strength and the spin-flip scattering strength and direction. We are also interested in the 0-$\pi$ transition and the second harmonic appearance, as well as the symmetries which occur in the current-phase relation, for different geometries of the two spin-flip interfaces. 
				\par In addition, we show how the Critical current of our junction is affected to the change of the above parameters and under which conditions it is optimized.
				\par Finally, we study the supercurrent flowing at zero phase difference (ZPC) of the two superconductors. We emphasize the conditions under which it is non-zero and also examine the cases it becomes maximized.
			\end{abstract}
			\maketitle
		\end{titlingpage}
	
	\tableofcontents

 \chapter{Introduction}
  In the beginning of this thesis, we introduce the reader to the basic theoretical concepts, in which our problem is based.We study the supercurrent of Josephson junctions, in which the region between the two superconductors is an intermediate layer of a metal with Rashba spin-orbit interaction, in the presence of barriers with magnetization on the interfaces. 
  
  \section{Theory of Superconductivity}
  
\par The phenomenon of superconductivity was discovered by Heike Kamerlingh Onnes in Netherlands, in 1911. He was the first to observe that, in certain material and in temperatures below a specific critical value T$_{c}$ , the electrical resistance becomes exactly zero \cite{bok}. After this first observation, hundreds of materials that become superconducting were discovered, with various critical temperatures. The highest temperature superconductor discovered ,until now,is the hydrogen sulfide (H$_{2}$S),  with T$_{c}$=203K, but at extremely high pressures \cite{hs}.
\par In 1933, Walther Meissner and Robert Ochsenfeld discovered that, inside every superconductor the magnetic field becomes zero: \textbf{B}=0 , making the superconductor a perfect diamagnet. This phenomenon, the expulsion of the magnetic field inside a superconductor, is called the Meissner effect. The explanation of this effect was given by the brothers Fritz and Heinz London, who proved that the magnetic field inside a superconductor has an exponential decay from the surface, with a decay length $\lambda$ , called  London penetration depth. In order for this to happen, the superconductor sets up electric currents on its surface, whose magnetic field cancels the applied magnetic field within the superconductor.
\par The phenomenon of superconductivity was theoretically explained in 1957, 46 years after it's discovery, by Bardeen, Cooper, and Schrieffer \cite{schr}. They developed the first microscopic theory of superconductivity, named the BCS theory, and received the Nobel Prize in Physics for it, in 1972. They proposed that the electrons of a superconductor that are near the Fermi surface attract indirectly through the crystal lattice, which is described as an exchange of phonons. This attraction overcomes the Coulomb repulsion and the electron form pairs, called Cooper pairs. Cooper pairs feel no scattering and so the supercurrent occurs. In T$>$T$_{c}$ though, the thermal vibration energy of the lattice becomes greater than the pairing energy of the electrons and so the Cooper pair breaks and there is no longer superconducting phenomenon.

\section{The Josephson Effect  }

\par In 1962 Brian David Josephson predicted theoretically that two superconductors that are coupled by a weak link, which link may consist of a normal metal ,an insulator, or a constriction that weakens the superconductivity in general, can still let the supercurrent flow through them \cite{JJ}.This macroscopic phenomenon was given the name Josephson effect. This process can be described as a quantum tunneling effect of the Cooper pair of electrons.

\bigskip
\begin{center}
	\includegraphics[scale=2]{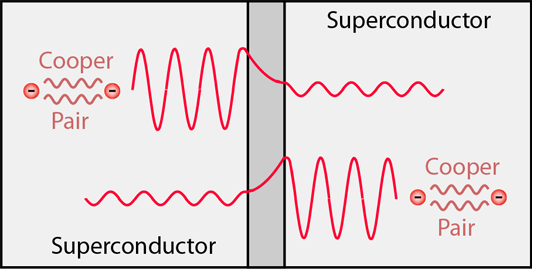}
	
	Figure 1.1: Tunneling effect of the Cooper pair through a weak link \cite{hyp}
\end{center} 

\newpage
\par Josephson proved that, for a short junction, the supercurrent, the current that flows through the junction when the voltage V=0, and the phase difference $\phi$ across the junction , which is the difference in the phase factor  between the order parameter of the two superconductors, are connected through the relation: 
	\begin{equation}
		I_{s}=I_{c}sin(\phi) 
	\end{equation}
	where I$_{c}$ is the supercurrent amplitude (with no external voltage applied) and $\phi$=$\chi$$_{2}$ -$\chi$$_{1}$, where $\chi$$_{i}$ is the phase of each superconductor.This phenomenon is known as the DC Josephson effect. There also exists the AC Josephson effect, but it will not concern us in this thesis.
	
\par The DC Josephson effect is described by a process known as Andreev reflection, named after Alexander F. Andreev \cite{andr}. This reflection is a particle scattering which occurs at the interfaces between the superconductor S and a normal metal N and explains how a normal current in the N side becomes a supercurrent in the S. Andreev proposed that an electron which approaches incidentally the interface from the N side can form a Cooper pair on the S side, and so it passes through the superconductor, with another electron with opposite momentum and spin and, at the same moment, reflect a hole inside the N region. 

 \bigskip
 
 \begin{center}
 	\includegraphics[scale=0.2]{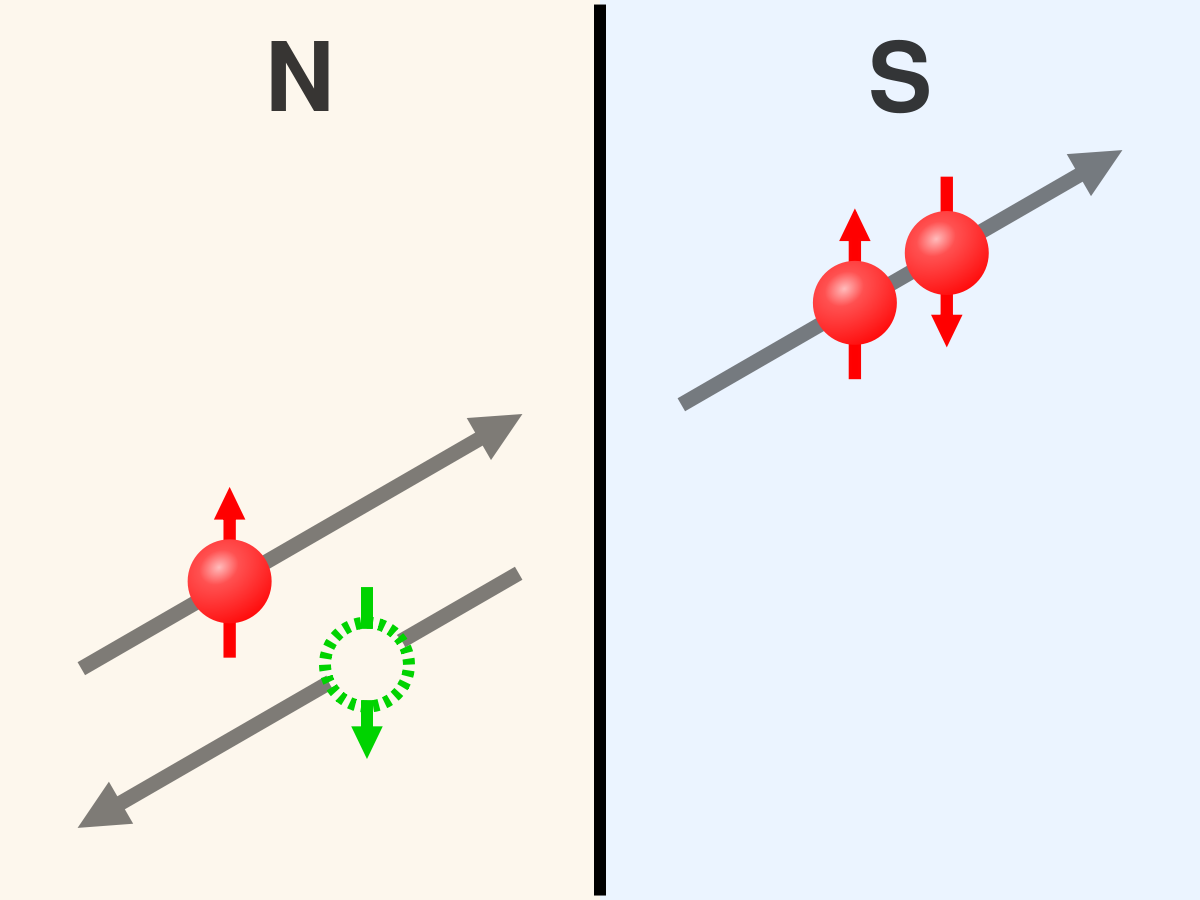}
 	
 	Figure 1.2:The Andreev reflection process schematically \cite{hypp}
 \end{center}

\newpage
\section{The two-dimensional electron gas (2DEG)  }

\par A two-dimensional electron gas (2DEG), is an electron gas that freely moves in two dimensions, while in the third dimension it is tightly confined and thus this dimension is ignored. 2DEGs are mostly found in semiconductor structures ,such as transistors. The type of 2DEG we study in this thesis, is a 2DEG which also has a Rashba spin-orbit coupling, also referred as the Rashba effect \cite{rshb}. This Rashba effect is described by the Rashba Hamiltonian: 	\begin{equation}
H_{R}=\alpha(\vec{\sigma} \times \vec{p})\cdot \hat{z}
\end{equation}
where $\alpha$ is the Rashba coupling,  $\vec{p}$  is the momentum and $\vec{\sigma}$ is the Pauli matrix vector, while $\hat{z}$ is the unit vector that is in the direction perpendicular to the two-dimensional plane.The study of spin-orbit coupling (SOC) and related effects
is one of the most active fields in mesoscopic physics.

\section{Previous work on the subject}

\par Since the discovery of the Josephson effect, there has been a continuously growing interest in the fundamental physics and applications of this effect. The achievements in Josephson-junction technology have made it possible to develop a variety of sensors for detecting ultralow magnetic fields and weak electromagnetic radiation, as well as the fabrication of ultrafast digital rapid single flux quantum (RSFQ) circuits \cite{lkr,dev}. Josephson junctions of different types have been studied, with the most important of them the superconductor-normal metal-superconductor (S/N/S) and the superconductor-ferromagnet-superconductor
(S/F/S) junctions, which we will discuss in the following, as well as the (S/2DEG/S) junction, which is studied in this thesis.
\par The understanding of the physics behind the Josephson effect was quite challenging. A.F.Andreev explained the phenomenon in 1964 establishing the concept of the so-called Andreev reflection, which includes another form of charge transport. A quasiparticle located in the weak link cannot penetrate directly from a normal metal into a superconductor if its energy is smaller than the superconducting energy gap. However, an electron with momentum k impinging on one of the interfaces can be converted into a hole moving in the opposite direction, generating a Cooper pair in
the superconductor. This hole is consequently Andreev reflected at the second interface and is converted back to an electron, leading to the destruction of a Cooper pair (Fig. 1.3). As a result of this cycle, a pair of correlated electrons is transferred from one superconductor to another, creating a supercurrent flow across the junction.
\begin{center}
	\includegraphics[scale=0.6]{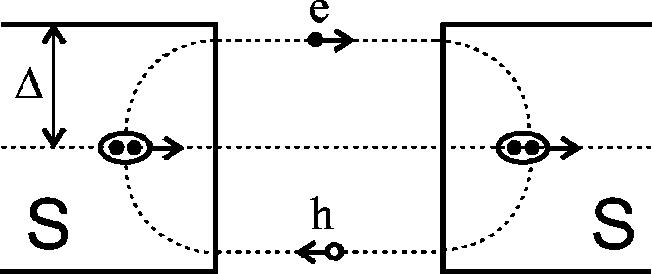}
	
	Figure 1.3:The cycle of the Andreev reflection \cite{snss}
\end{center}

\par The Andreev reflection amplitudes depend on the corresponding quasiparticles' phases $\chi_{1,2}$ and so, the resulting current depends on the phase difference $\phi=\chi_{1}-\chi_{2}$ of the two superconductors. Due to the electron-hole interference in the quantum well, formed by the pairing potentials of the superconducting electrodes, current carrying standing waves with quantized energy appear in the weak-link region The corresponding quantum states are referred to as Andreev bound states, which has been studied extensively \cite{klk}. 

\begin{center}
	\includegraphics[scale=0.6]{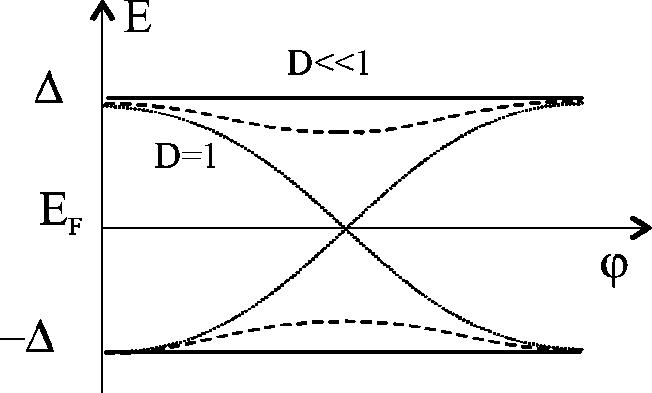}
	
	\begin{flushleft}
		Figure 1.4:Energy-phase relation for Andreev bound states in a short weak link: Tunneling limit (low transparency, solid lines), ballistic regime (dotted line), intermediate case (dashed lines) \cite{snss}.
	\end{flushleft}
\end{center}

\par In the general case, the supercurrent I($\chi$), which is the sum of the partial currents transported via Andreev bound states, can be decomposed into a Fourier series\cite{tnk}: 
\begin{equation}
I(\phi)=\pd{E(\phi)}{\phi}=\sum_{n\geq1}(I_{n}sin(n\phi)+J_{n}cos(n\phi))
\end{equation}
where I$_{n}$ and J$_{n}$ are coefficients to be determined. Also, J$_{n}$ vanishes if time-reversal symmetry is not broken. 
\par An interesting case is the junction with $I_{c}<0$, which is called the $\pi$-junction \cite{blv,sp1,sp2,sp3}. Such a junction has an energy minimum at $\chi=\pi$, providing a phase shift of $\pi$ in the ground state. $\pi$-junction may be used as a phase inverter in superconducting digital circuits (SQUIDS) \cite{trz,qb}, and are proposed as candidates for engineering a quantum two-level system, or qubit, which is the basic element of a quantum computer.
\par A lot of study has been done for the S/N/S junction (see for example \cite{bj,snss}). In the clean limit, in which the mean free path of an electron is larger than the distance between the superconducting electrodes, only a small number of electrons from the S metals moving almost perpendicular to the SN interface could penetrate from S into N and finally from N into the second electrode, thus providing a Josephson coupling in the structure. In these junctions, the current-phase relation transforms from the sinusoidal form as T$\rightarrow$Tc, to a saw-toothed curve, at low T.
\par There is also a continuously growing interest in charge and spin transport in contacts between superconductors and ferromagnets (S/F/S junction) (see \cite{frst,brev}). The most important feature of these junction was the ability to create a crossover from the 0 to $\pi$ state (0-$\pi$ transition). The first experimental observations of this effect were reported by Ryazanov et al., in 2000 \cite{rzN}. In an SF bilayer correlated electrons and holes, having opposite spin directions, are under the exchange field of a ferromagnet. This results in an energy shift between these quasiparticles and the creation of a nonzero momentum Q of Cooper pairs. As a result, the amplitude of superconducting correlations oscillates spatially in the ferromagnetic metal as cos(Qx) \cite{zp}. The sign change of this amplitude is equivalent to periodic 0-$\pi$ phase jumps for certain lengths of the ferromagnet. This transition can be achieved either by decreasing the temperature \cite{rzN}, or by changing the magnetization directions of the layers \cite{flt}.
\par For low-temperature electronics it is quite natural to combine the technology of modern superconductors with that of semiconductors. One of the practical realizations of such a combined approach is to couple two superconducting electrodes by a 2D electron gas (S/2DEG/S junction) \cite{MRS}. In addition, it was shown experimentally that the strength of the Rashba coupling can be controlled by a gate voltage \cite{NT,gv}. In this kind of junctions, the combined effects of topology and electronic correlations can be used to investigate unconventional Josephson effects, such as the anomalous Josephson effect ($I(\phi=0)\ne0$), which we will also examine in the following chapters. In this case, the corresponding ground state of the junction is found at a phase $\phi_{0}\ne\{0,\pi\}$, where the Josephson current is zero. In addition, for a strong Rashba interaction, the projection of electron spin is strongly correlated with the direction of motion, and so, left- and right-moving electrons with the same energy always have opposite spin projections. This fact leads to the phenomenon of direction-dependent critical current. Also, the combined effect of Rashba and spin-flip interactions \cite{sp4,sp5} leads to phenomena which would otherwise require a magnetic field that is too large to be sustained by the superconducting leads, in an S/F/S junction \cite{MNt}.

\section{Organization of Thesis} 

\par In chapter 2 we will approach the Josephson junction problem as a 2-D quantum scattering system and solve the problem analytically, making use of the Bogoliubov-de Gennes (BdG) equations \cite{BdG}. We will construct the matching matrix from the boundary conditions of the problem and, by calculating it's determinant, we will calculate the supercurrent with a method named the
Furusaki-Tsukada method \cite{FTM}. This method finds the supercurrent by summing analytically all the contributions from
the Andreev reflection.  
\par In chapter 3, we will study our junction numerically. What will mostly concern us in this thesis is the current-phase relation as a function of the spin-orbit interaction on the 2DEG and the magnetization of the interfaces. 
\par Following, in chapters 4 and 5, we will study how the Critical current and the Zero-phase current, respectively, differs, for different geometries and spin-orbit amplitudes.
\par Finally, in the conclusion section, we will summarize our most significant results and we will draw some conclusions. We will also discuss further expansion of this Thesis subject.
\newpage
\chapter{Analytical approach of the problem}
To begin with, we consider a clean 2-D ballistic S/2DEG/S Josephson junction, as shown in figure 2.1 below. It consists of two identical superconductors on the sides, interrupted by a 2DEG with Rashba spin-orbit coupling. In the region between each superconductor and the 2DEG exist very thin interface barriers at x=0 and x=d (the black regions referred as IF in the figure). The barriers are in the z-y plane and have both normal
scattering and spin-flip effect, due to their magnetization. 
\begin{center}
\includegraphics[scale=0.7]{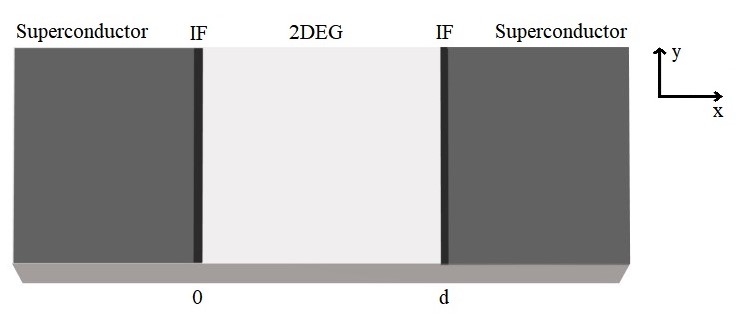}

Figure 2.1:Schematic of the S/2DEG/S junction
\end{center}

\newpage
\section{ Bogoliubov-de Gennes equations}

\par BdG equations are a generalization of the BCS theory, in which space dependent quantities are introduced. At first, they introduce a space dependent pairing potential ,for the Cooper pair of electrons: $\Delta=\Delta(r)$. Next, they use a "mean field" approximation to describe, finally, the "Schroedinger" equation of the components $u(r)$ and $v(r)$ of a spinor $\psi(r)$, which are the amplitudes of the electron and hole part of the quasiparticle excitations correspondingly:  
\begin{equation}
\vec{\Psi}(\vec{r})=\begin{pmatrix}
u(\vec{r}) \\ v(\vec{r})
\end{pmatrix}
\end{equation}
The effective Hamiltonian that acts on this spinor has the form of a matrix:
\begin{equation}
H=\begin{pmatrix}
\hat{H_{0}}&\hat{\Delta}(\vec{r}) \\ \hat{\Delta}^*(\vec{r})&-\hat{H_{0}^*}
\end{pmatrix}
\end{equation}
The pairing potential function of our problem is:
\begin{equation}
\hat{\Delta}(x)=
\begin{cases}
i \sigma_{y}\Delta , & \text{$x\in(-\infty,0)$} \\
0, & \text{$x\in[0,d]$} \\
i \sigma_{y}\Delta e^{i\phi}, & \text{$x\in(d,\infty)$}
\end{cases}
\end{equation}
$\hat{H_{0}}$ is the normal Hamiltonian, which in our problem has the form:
\begin{equation}
\hat{H_{0}}=-\hbar^2\dfrac{d}{dx}\dfrac{1}{2m(x)}\dfrac{d}{dx}+U(x)+\hat{H}_{RSOC}-\mu
\end{equation}
with the first term being the kinetic energy, m(x) the effective mass function and $\mu$ the chemical
potential.U(x) is the interface potential:
\begin{equation}
U(x)=[U_{L}-\hat{\vec{\sigma}}\cdotp \vec{N}_{L}]\delta(x)+[U_{R}-\hat{\vec{\sigma}}\cdotp \vec{N}_{R}]\delta(x-d)
\end{equation}
In the equation above, $U_{\nu}$ represents the normal scattering strength and $\vec{N_{\nu}}$ represents the spin-flip scattering strength ($\nu$=L,R), which is proportional to the magnetization of the interfaces, while $\vec{\sigma}$ is the Pauli matrices vector.
\par The term $\hat{H}_{RSOC}$ describes the Rashba spin-orbit coupling (RSOC) in the 2DEG region and is given by the relation:
\begin{equation}
\hat{H}_{RSOC}=\dfrac{\lambda_{0}}{\hbar}(\sigma_{y}p_{x}-\sigma_{x}p_{y})[\theta(x)-\theta(x-d)]
\end{equation}
$\lambda_{0}$ is called RSOC constant and defines the amplitude of spin-orbit coupling, $\sigma_{i}$ and $p_{i}$ are the Pauli matrix and momentum vector components, respectively, in the i direction. Lastly, $\theta$(x) is the step function.

\section{Eigenfunctions of the  junction's layers}

\par As mentioned before, in order to calculate the supercurrent of our junction, we have to use the method of Furusaki-
Tsukada. To do that, we have to write the wavefunction in each layer, then apply the boundary conditions on the interfaces and finally construct the matching matrix from them. 
\par We study the problem using the short range potential scattering theory. Thus, we solve the BdG equations separately in each region and match our solutions at the respective boundaries. In order to do that, we first have to find the eigenfunctions of each layer.
\bigskip
\begin{flushleft}
	\textbf{{\large 2.2.1 Eigenfunctions of the superconductor}}
\end{flushleft}
\par Our junctions consists of two superconductors. The left (L) superconductor extends for $x\in(-\infty,0)$, while the right (R) extends for $x\in(d,\infty)$. Thus, the planar wave on the L is described by $e^{(-ik_{eL}x)}$ for electrons and by $e^{(+ik_{hL}x)}$ for the holes. On the R we have the opposite signs, for the transmitted wavefunctions: $e^{(+ik_{eR}x)}$ for electrons and $e^{(-ik_{hR}x)}$ for holes. Our system though is in two dimensions. Because of the translational invariance along the y axis, the momentum in that direction, parallel to the interface, $k_{p}$, is conserved. Thus, the wavefunction in the y direction will by described by $e^{ik_{p}y}$.
\par We use the index $\nu$=L,R for left and right superconductor respectively. So, for each superconductor, the total wavefunction will be a linear combination of the eigenfunctions below:
\bigskip
\begin{equation}
\vec{\Psi_{e\uparrow\nu}}=\begin{pmatrix}
u_{\nu}(E)e^{i\frac{\chi_{\nu}}{2}}\\0\\0\\v_{\nu}(E)e^{-i\frac{\chi_{\nu}}{2}}
\end{pmatrix}e^{\pm ik_{e\nu}x}e^{ik_{p}y}
\end{equation} 
\begin{equation}
\vec{\Psi_{e\downarrow\nu}}=\begin{pmatrix}
0\\u_{\nu}(E)e^{i\frac{\chi_{\nu}}{2}}\\-v_{\nu}(E)e^{-i\frac{\chi_{\nu}}{2}}\\0
\end{pmatrix}e^{\pm ik_{e\nu}x}e^{ik_{p}y}
\end{equation}
\begin{equation}
\vec{\Psi_{h\uparrow\nu}}=\begin{pmatrix}
0\\v_{\nu}(E)e^{i\frac{\chi_{\nu}}{2}}\\-u_{\nu}(E)e^{-i\frac{\chi_{\nu}}{2}}\\0
\end{pmatrix}e^{\pm ik_{h\nu}x}e^{ik_{p}y}
\end{equation}
\begin{equation}
\vec{\Psi_{h\downarrow\nu}}=\begin{pmatrix}
v_{\nu}(E)e^{i\frac{\chi_{\nu}}{2}}\\0\\0\\u_{\nu}(E)e^{-i\frac{\chi_{\nu}}{2}}
\end{pmatrix}e^{\pm ik_{h\nu}x}e^{ik_{p}y}
\end{equation}
\bigskip

In the relations above, E is the energy of the wavefunction and $\chi_{\nu}$ is the phase of the superconducting order parameter. In addition, $k_{q\nu}$ (where q stands for the quasiparticle type: q=e,h) are the wavenumbers and are given from the relation: 
\begin{equation}
k_{q\nu}=\sqrt{\dfrac{2m}{\hbar^2}(\pm \Omega_{\nu} +\mu-U)-k_{p}^2}
\end{equation}
where $\Omega_{\nu}=sgn(E)\sqrt{E^2-\abs{\Delta_{\nu}}^2}$ and the symbol "sgn" stands for the signature function. The coherence factors, u and v, of the superconductors, are:
\begin{equation}
u_{\nu}(E)=\sqrt{\dfrac{1}{2}(1+\dfrac{\Omega_{\nu}}{E})}
\end{equation}
\begin{equation}
v_{\nu}(E)=sgn(E)\sqrt{\dfrac{1}{2}(1-\dfrac{\Omega_{\nu}}{E})}
\end{equation}
\bigskip
\begin{flushleft}
	\textbf{{\large 2.2.2 Eigenfunctions of the 2DEG}}
\end{flushleft}
\par As mentioned before, the Hamiltonian of the 2DEG is of the form:
\begin{equation}
\hat{H}_{RSOC}=\dfrac{\lambda_{0}}{\hbar}(\sigma_{y}p_{x}-\sigma_{x}p_{y})
\end{equation}
\bigskip

The eigenfunctions of this Hamiltonian are:
\begin{equation}
\vec{\Psi_{e1}}=\dfrac{1}{\sqrt{2}}\begin{pmatrix}
e^{\pm i\chi(a1)}\\1\\0\\0
\end{pmatrix}e^{\pm iq_{e1}x}e^{ik_{p}y}
\end{equation} 
\begin{equation}
\vec{\Psi_{e2}}=\dfrac{1}{\sqrt{2}}\begin{pmatrix}
-e^{\pm i\chi(a2)}\\1\\0\\0
\end{pmatrix}e^{\pm iq_{e2}x}e^{ik_{p}y}
\end{equation} 
\begin{equation}
\vec{\Psi_{h1}}=\dfrac{1}{\sqrt{2}}\begin{pmatrix}
0\\0\\-e^{\pm i\chi(b1)}\\1
\end{pmatrix}e^{\mp iq_{h1}x}e^{ik_{p}y}
\end{equation}
\begin{equation}
\vec{\Psi_{h2}}=\dfrac{1}{\sqrt{2}}\begin{pmatrix}
0\\0\\e^{\pm i\chi(b2)}\\1
\end{pmatrix}e^{\mp iq_{h2}x}e^{ik_{p}y}
\end{equation}
\bigskip

In the relations above, the upper sign is for particle moving towards the R direction and the lower is for the L direction, and $q_{pi}$, with p=e,h being the particle index and and i=1,2 being the spin-mode index, are the wavevectors of the particles.We also define $a_{i}$ and $b_{i}$, as the angles between the particle wavevectors and the x axis:
\newpage
\begin{equation}
a_{i}=\arctan{\dfrac{k_{p}}{q_{ei}}}
\end{equation}
\begin{equation}
b_{i}=\arctan{\dfrac{k_{p}}{q_{hi}}}
\end{equation}
The coefficients $\chi(a_{i})$ and $\chi(b_{i})$ are the angles between the particle wavevectors and the y axis:
\begin{equation}
\chi(a_{i})=\dfrac{\pi}{2}-a_{i}
\end{equation}
\begin{equation}
\chi(b_{i})=\dfrac{\pi}{2}-b_{i}
\end{equation}
\bigskip

Now, to define the particle wavevectors, we define $k_{F}$ and $k_{R}$ as the Fermi and Rashba wavevector, respectively, given from the relations:
\begin{equation}
k_{F}=\sqrt{\dfrac{2mE_{F}}{\hbar^2}}
\end{equation}
\begin{equation}
k_{R}=\dfrac{m\lambda_{0}}{\hbar^2}
\end{equation}
The wavevectors of the electrons are:
\begin{equation}
q_{ei}=\sqrt{(\sqrt{k^2_{R}+k^2_{F}+\dfrac{2mE}{\hbar^2}}\mp k_{R})^2-k^2_{p}}
\end{equation}
with the upper sign for i=1 and the lower for i=2.
In a similar way, the wavevectors of the holes are:
\begin{equation}
q_{hi}=\sqrt{(\sqrt{k^2_{R}+k^2_{F}-\dfrac{2mE}{\hbar^2}}\mp k_{R})^2-k^2_{p}}
\end{equation}
\par So, for the 2DEG, the total wavefunction will be a lineal combination of all the 8 eigenfunctions written above(relations (2.14)-(2.17). 
\bigskip

\begin{center}
	\includegraphics[scale=0.6]{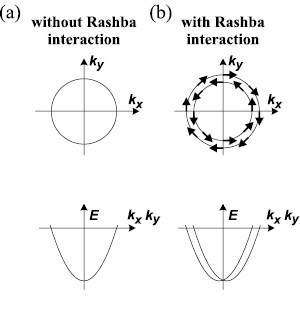}
	
	Figure 2.2: Schematic of the Rashba effect on the k-space and the Energy
\end{center} 

\section{Constructing the matching matrix from the boundary conditions}

\par In this last section of the chapter, we write down the boundary-condition equations in the form of a matrix. This will be the final step in the process of computing the supercurrent.
\bigskip

\textbf{\begin{flushleft}
		{\large 2.3.1 The boundary-condition equations}
\end{flushleft}}
In section 2.2 we found the eigenfunctions in each layer. As mentioned before, the wavefunction of the layer is a linear combination of these wavefunctions. Thus, our whole wavefunction has 16 unknown coefficients: 4 for each superconductor and 8 for the intermediate layer. In order to define those coefficients we have to solve the boundary conditions for each interface. So we have:
\bigskip

\begin{equation}
\Psi(x)|_{x=x_{i}^+}=\Psi(x)|_{x=x_{i}^-}
\end{equation}
\begin{equation}
\hat{v}\Psi(x)|_{x=x_{i}^+}=\hat{v}\Psi(x)|_{x=x_{i}^-}+\hat{U}\Psi(x_{i})
\end{equation}
\bigskip

where i=1,2 and $x_{1}$=0 for left interface and $x_{2}=d$ for the right. We also make use of the velocity operator:
\bigskip

\begin{equation}
\hat{v}=\begin{pmatrix}
-\dfrac{\hbar}{m(x)}\dfrac{d}{dx}&\dfrac{\lambda_{0}}{\hbar}\Theta(x)&0&0\\-\dfrac{\lambda_{0}}{\hbar}\Theta(x)&-\dfrac{\hbar}{m(x)}\dfrac{d}{dx}&0&0\\0&0&\dfrac{\hbar}{m(x)}\dfrac{d}{dx}&-\dfrac{\lambda_{0}}{\hbar}\Theta(x)\\0&0&\dfrac{\lambda_{0}}{\hbar}\Theta(x)&\dfrac{\hbar}{m(x)}\dfrac{d}{dx}
\end{pmatrix}
\end{equation}
\bigskip

in which $\Theta(x)$=$\theta(x)-\theta(x-d)$.
\bigskip 

Finally, we define the scattering operator $\hat{U}$ as:
\begin{equation}
\hat{U}=\begin{pmatrix}
\hat{Z}_{\nu}&\hat{0}\\\hat{0}&\hat{Z}^*_{\nu}
\end{pmatrix}\begin{pmatrix}
1&0&0&0\\0&1&0&0\\0&0&-1&0\\0&0&0&-1
\end{pmatrix}
\end{equation}
In the equation above, $\hat{0}$ is a 2x2 matrix with zeroes and $\hat{Z_{\nu}}$ is also a 2x2 matrix of the form:
\begin{equation}
\hat{Z_{\nu}}=\dfrac{2m(x)}{\hbar^2}[U_{\nu}\hat{I}-\hat{\vec{\sigma}}\cdotp\ \vec{N}_{\nu}]
\end{equation}
with $U_{\nu}$ and $\hat{N}_{\nu}$ from (2.5).
\newpage

\textbf{\begin{flushleft}
	{\large 	2.3.2 The matching matrix}
\end{flushleft}}
Now, we are finally able to construct the matrix. The above equations can be written as:
\bigskip

\begin{equation}
\begin{pmatrix}
L_{1}&S_{1}&0&0\\L_{2}&0&S_{2}&0\\0&S'_{1}&0&R_{1}\\0&0&S'_{2}&R_{2}
\end{pmatrix}\begin{pmatrix}
\alpha\\\beta\\\gamma
\end{pmatrix}=B_{ps}
\end{equation}
\bigskip

where $L_{i}$, $S_{i}$, $S'_{i}$, $R_{i}$ (i=1,2) , as well as the elements 0, are all 4x4 matrices. In addition, $\alpha$ and $\gamma$ are 4-component columns, with elements the reflection $\alpha_{ps}$ and transmission $\gamma_{ps}$ coefficients respectively, while $\beta$ is an 8-component column, with elements $\beta_{psi}$. The index p is referred to the particle, s to spin and i to the direction (right going and left going wave). To make the columns, we take all the possible combinations of the above indexes. Also, $B_{ps}$ is a column vector with 16 components that describes the incident wave.
\par Now, to define the block matrices, we use the symbol  $\hat{1}$ for the unitary 2x2 matrix and $\hat{1}_{a}$ for the matrix below:
\begin{equation}
\hat{1}_{a}=\begin{pmatrix}
0&-1\\1&0
\end{pmatrix}
\end{equation}
\par Also. in order to write the matrices below, we have normalized our parameters. All the lengths are normalized on the Fermi wavevector $ k^{-1}_{F}$ and all the energy units over the Fermi energy $E_{F}$. We also use the symbol $m^*$ for the normalized (effective) mass on the 2DEG over the mass of the particles on the superconductors, and $\lambda$, which is the normalized spin-orbit coupling constant:
\begin{equation}
\lambda=\dfrac{\lambda_{0}k_{F}}{E_{F}}
\end{equation}
Finally, the current units are normalized over $I_{n}$:
\begin{equation}
I_{n}=\dfrac{e\Delta}{\hbar}
\end{equation}
where $\Delta$ is the pair potential, given in (2.3).
\bigskip

So, $L_{1}$ and $L_{2}$ are the 4x4 matrices:
\begin{equation}
L_{1}=\begin{pmatrix}
u_{L}\hat{1}&v_{L}\hat{1}\\(-ik_{eL}\hat{1}-Z_{L})u_{L}&(ik_{hL}\hat{1}-Z_{L})v_{L}
\end{pmatrix}e^{i\frac{\chi _{L}}{2}}
\end{equation}
\begin{equation}
L_{2}=\begin{pmatrix}
v_{L}\hat{1}_{a}&u_{L}\hat{1}_{a}\\(-ik_{eL}\hat{1}-Z^*_{L})\hat{1}_{a}v_{L}&(ik_{hL}\hat{1}-Z^*_{L})\hat{1}_{a}u_{L}
\end{pmatrix}e^{-i\frac{\chi _{L}}{2}}
\end{equation}
\bigskip

In the matrices above, $k_{q\nu}$, $u_{\nu}$ and $v_{\nu}$ are given from the relations (2.11),(2.12) and (2.13) respectively.
Now, before we enter the 2DEG area, we first have to define the following spinors:
\bigskip

\begin{equation}
e_{1\nu}=\begin{pmatrix}
\mp ie^{\pm a_{1}}\\ 1
\end{pmatrix},
e_{2\nu}=\begin{pmatrix}
\pm ie^{\pm a_{2}}\\ 1
\end{pmatrix}
\end{equation}
\begin{equation}
h_{1\nu}=\begin{pmatrix}
\pm ie^{\pm b_{1}}\\ 1
\end{pmatrix},
h_{2\nu}=\begin{pmatrix}
\mp ie^{\pm b_{2}}\\ 1
\end{pmatrix}
\end{equation}
\bigskip

where $\nu$=L,R as always, and the upper sign is for $\nu$=L. Also, $a_{i}$ and $b_{i}$ are the angles defined from the relations (2.18) and (2.19). Next, the matrices $S_{1}$ and $S_{2}$ are:
\bigskip
\begin{equation}
S_{1}=\begin{pmatrix}
E_{1}&E_{2}
\end{pmatrix}
\end{equation}
\begin{equation}
S_{2}=\begin{pmatrix}
H_{1}&H_{2}
\end{pmatrix}
\end{equation}

Where :
\bigskip

\begin{equation}
E_{i}=\begin{pmatrix}
-e_{iR}&-e_{iL}\\(-\dfrac{iq_{ei}}{m^*}\hat{1}-\lambda\hat{1}_{a})e_{iR}&(\dfrac{iq_{ei}}{m^*}\hat{1}-\lambda\hat{1}_{a})e_{iL}
\end{pmatrix}
\end{equation}
\begin{equation}
H_{i}=\begin{pmatrix}
-h_{iR}&-h_{iL}\\(\dfrac{iq_{hi}}{m^*}\hat{1}-\lambda\hat{1}_{a})h_{iR}&(-\dfrac{iq_{hi}}{m^*}\hat{1}-\lambda\hat{1}_{a})h_{iL}
\end{pmatrix}
\end{equation}

Above, $q_{ei}$ and $q_{hi}$ are given in (2.25) and (2.26), (i=1,2).
\bigskip

The matrices $S'_{1}$ and $S'_{2}$ are:
\bigskip
\begin{equation}
S'_{1}=\begin{pmatrix}
E'_{1}&E'_{2}
\end{pmatrix}
\end{equation}
\begin{equation}
S'_{2}=\begin{pmatrix}
H'_{1}&H'_{2}
\end{pmatrix}
\end{equation}
Where :
\bigskip

\begin{equation}
E'_{i}=\begin{pmatrix}
e_{iR}e^{iq_{ei}d}&e_{iL}e^{-iq_{ei}d}\\(\dfrac{iq_{ei}}{m^*}\hat{1}+\lambda\hat{1}_{a})e_{iR}e^{iq_{ei}d}&(-\dfrac{iq_{ei}}{m^*}\hat{1}+\lambda\hat{1}_{a})e_{iL}e^{-iq_{ei}d}
\end{pmatrix}
\end{equation}
\begin{equation}
H'_{i}=\begin{pmatrix}
h_{iR}e^{-iq_{hi}d}&h_{iL}e^{iq_{hi}d}\\(-\dfrac{iq_{hi}}{m^*}\hat{1}+\lambda\hat{1}_{a})h_{iR}e^{-iq_{hi}d}&(\dfrac{iq_{hi}}{m^*}\hat{1}+\lambda\hat{1}_{a})h_{iL}e^{iq_{hi}d}
\end{pmatrix}
\end{equation}
\bigskip

Finally, the matrices $R_{1}$ and $R_{2}$ are:
\bigskip

\begin{equation}
R_{1}=-\begin{pmatrix}
u_{R}\hat{1}&v_{R}\hat{1}\\(ik_{eR}\hat{1}-Z_{R})u_{R}&(-ik_{hR}\hat{1}-Z_{R})v_{R}
\end{pmatrix}e^{i\frac{\chi _{R}}{2}}
\end{equation}
\begin{equation}
R_{2}=-\begin{pmatrix}
v_{R}\hat{1}_{a}&u_{R}\hat{1}_{a}\\(ik_{eR}\hat{1}-Z_{R})v_{R}\hat{1}_{a}&(-ik_{hR}\hat{1}-Z_{R})u_{R}\hat{1}_{a}
\end{pmatrix}e^{-i\frac{\chi _{R}}{2}}
\end{equation}
\newpage
\section{The Supercurrent}

\par Now that we have the matrix of the equation (2.31), we find it's determinant $\Gamma$. $\Gamma$ is a function of the phase $\chi$=$\phi_{R}$-$\phi_{L}$, the energy E and the parallel to the interface momentum $k_{p}$:
\begin{equation}
\Gamma=\Gamma(\chi,E,k_{p})
\end{equation}
\par Next, we set $\phi_{L}$=0 and so $\chi$=$\phi_{R}$. Because we are working for a non-zero temperature T, we have thermal excitations. Thus, we have to use the Matsubara frequency summation method. This method implies that we use discrete imaginary frequencies $i\omega_{n}$, instead of the energy E, and sum over them to find the thermal excitations' contribution. 
\bigskip

So, now,for the supercurrent I we have:
\begin{equation}
I=-\dfrac{e}{\hbar\beta}\sum_{\omega_{n}}\sum_{k_{p}}\dfrac{1}{\Gamma}\dfrac{\partial\Gamma}{\partial\chi}
\end{equation}

We now can use the method of Laplace expansion, which makes us find the exact dependence of $\Gamma$ from the phase $\chi$ and so we have:

\begin{equation}
\Gamma(\chi,\omega_{n},k_{p})=A(\omega_{n},k_{p})cos(2\chi)+B(\omega_{n},k_{p})sin(\chi)+C(\omega_{n},k_{p})cos(\chi)+D(\omega_{n},k_{p})
\end{equation}

Thus, the final and most crucial formula, which is the one we use in the next section is:
\bigskip 

\begin{equation}
I=\dfrac{e}{\hbar\beta}\sum_{\omega_{n}}\sum_{k_{p}}\dfrac{2Asin(2\chi)-Bcos(\chi)+Csin(\chi)}{Acos(2\chi)+Bsin(\chi)+Ccos(\chi)+D}
\end{equation}
\bigskip

\newpage
\chapter{Supercurrent in a S/2DEG/S junction}
\par In this chapter we solve numerically the equations of chapter 2 and then we examine the dependence of the supercurrent from the phase $\chi$. At first we study how these relations transform under the change of our other parameters separately. Later on, we focus our study on the (drastic) change of these relations for different interface magnetization geometries.

\section{General study of the current-phase \\ relation }

\par In this first section of the chapter, we study the general properties of the S/2DEG/S junction. First, we shall observe how the spin-orbit coupling itself affects the current-phase relation. This is shown in figure 3.1 below. In this figure, we display on the y axis the normalized current over $I_{n}$ (relation 2.35). In the x axis is the phase $\chi$ over $\pi$. In addition, for this graph all scattering interfaces are set zero (Z=$Z_{n}$=$Z_{m}$=0) and T=0.1$T_{c}$. We represent each graph by writing it's title on the legend and, next to it inside a parenthesis, the parameters. 
\par \textit{Comment}: in most of our graphs, we have $m^*$=1 and T=0.1$T_{c}$. So, these parameters are not refereed in the description of our figures. In the rare case we set $m^*$$\ne$1 or T$\ne$0.1$T_{c}$ we write it down, both before and in the description of the figure. Also, in the cases that $Z_{n}$=$Z_{m}$, we use the symbol Z to describe both of them.

\begin{center}
	\includegraphics[scale=0.5]{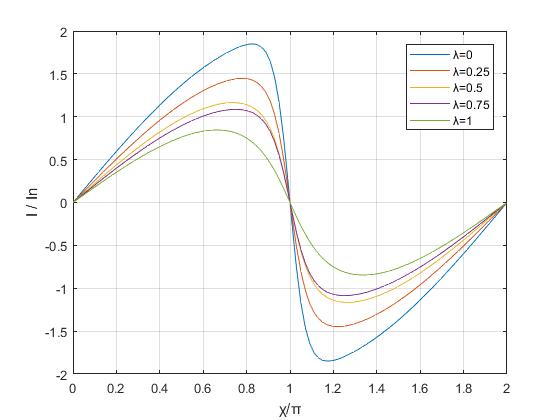}
	
Figure 3.1: Current-Phase Relation for various $\lambda$ ($k_{F}d$=10, Z=0)
\end{center}

 One can observe that the presence of $\lambda$ makes the curve smoother, more sinusoidal and also decreases it's amplitude, as it increases the misfit at the interfaces.
\par Next, we shall see the effect of the temperature (figure 3.2). As we know, the rise of the temperature decreases the gap of the superconductor.  
\begin{center}
	  \includegraphics[scale=0.5]{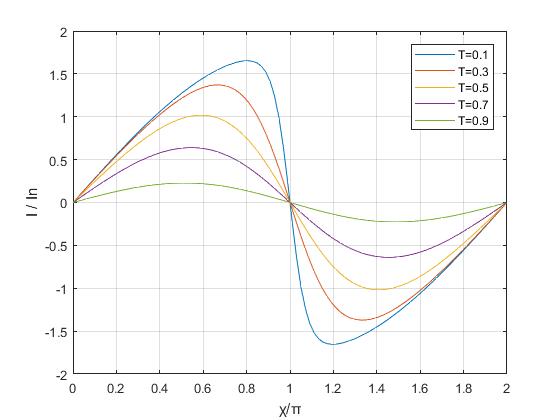}
	
	Figure 3.2: Current-Phase Relation for various $\dfrac{T}{T_{c}}$ ($k_{F}d$=10, $\lambda$=0.1, Z=0 )
\end{center}

\bigskip

By increasing the temperature, the current decreases drastically and tends to zero as we approach the value $T_{c}$.
\par Next up, we shall examine the effect of the scattering barriers. In figure (3.3)  we have a junction which includes barriers with increasing normal scattering effect (but not spin-flip):
\begin{center}
	 \includegraphics[scale=0.5]{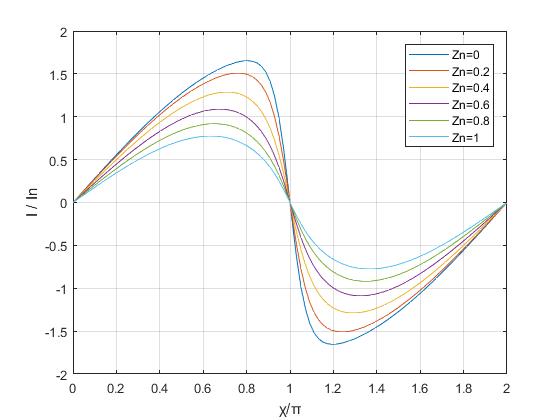}
	 
	Figure 3.3: Current-Phase Relation for various $Z_{n}$\\ ($k_{F}d$=10, $\lambda$=0.1, $Z_{m}=0$)
\end{center}

\bigskip

As expected, the normal scattering also makes our curve smoother and more sinusoidal, decreasing the current. Generally speaking, the position of the critical current tends to $\chi$=$\dfrac{\pi}{2}$, as the reflection on the interfaces becomes stronger. This happens because the interface reflection creates "misfit", and thus our junction becomes more realistic.
\par Another parameter that also creates "misfit" is the difference of the effective mass ($m^*$), between the 2DEG and the superconductor. As mentioned before, 2DEGs are found in semiconductor structures. In semiconductors commonly the effective mass becomes smaller. The dependence of the supercurrent from this $m*$ decrease is shown in figure (3.4). As $m*$ decreases, the misfit becomes rapidly larger.
 \begin{center}
 	\includegraphics[scale=0.5]{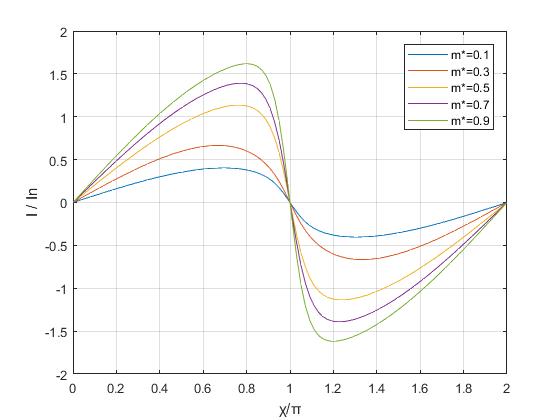}
 	
 	Figure 3.4: Current-Phase Relation for various values of $m^*$\\($k_{F}d$=10, $\lambda$=0.1, Z=0 )
 \end{center}

\par To end this section, we will examine the dependence of the supercurrent from the normalized length $k_{F}d$ of the 2DEG, which is shown in figure (3.5):
 \begin{center}
	\includegraphics[scale=0.5]{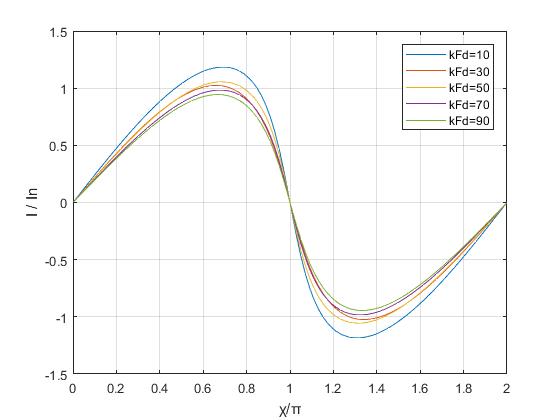}
	
	Figure 3.5: Current-Phase Relation for various values of $k_{F}d$ \\($\lambda$=0.1, $Z_n$=0.5, $Z_m$=0 )
\end{center}

\bigskip
\par In the graph above we have applied a scattering barrier without spin-flip. One can see that the current's amplitude slowly decays for longer junctions, without changing it's form. Such a slow decay is expected, as long as the 2DEG's length is quite smaller than the correlation length: $d<<\xi_{0}$.
\bigskip

\section{The interface magnetization effects I: \\The second harmonic}

\par In this section we examine separately how the spin-flipping effect, which appears due to the interfaces' magnetization, changes the current-phase relation we saw before. We will see that, contrary to the previous cases, the spin-flip changes dramatically, not only the amplitude, but also the scheme of this relation's graph. Also, the effect depends on the direction of the two magnetizations and the angle between them. Thus, we have a variety of interesting phenomena including the $\pi$-junction and the Zero-phase current, which we will discuss later.
\par To begin with, we plot the zero-phase relation with the magnetizations being in either the X or the Z axis. So, we have the 4 curves seen in figure (3.6). In the legend, the first letter is refereed to the direction of the magnetization of the left interface and the second for the right.

\begin{figure}
	\centering
	\includegraphics[scale=0.5]{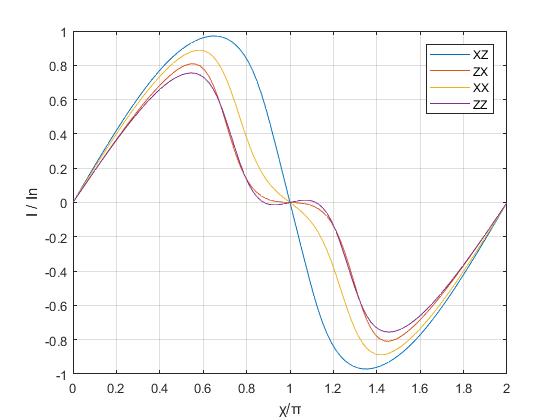}
	
	Figure 3.6: Current-Phase Relation for various magnetization geometries \\($k_{F}d$=10, $\lambda$=0.1, Z=0.5) 
\end{figure}

\par The most important feature of this graph is that, while for the XZ geometry we have a sine-like curve, in the other 3 geometries our curve is no longer sinusoidal but a second harmonic makes it's appearance. One can imagine that, as this second harmonic dominates the first, we have the process of the 0-$\pi$ transition, which is the transition of a sine-like curve into a (-sine)-like one. As we will see in the following, the second harmonic can also occur for a XZ geometry, but for different values of the parameters.
\par In the second harmonic, the transfer of Cooper pairs through the junction involves co-tunneling processes, which is the process of simultaneous tunneling of two or more electrons. A lot of research has been made in understanding this complex quantum-state transition and make it tunable experimentally.
\par This transition's process includes two steps, which we present in the next two figures. The first step is shown in figure (3.7) and is the domination of the second harmonic over the 0-junction. In this graph we achieve that by changing the scattering (both normal and magnetic) interfaces' amplitude, for the ZZ geometry:

\begin{center}
	\includegraphics[scale=0.5]{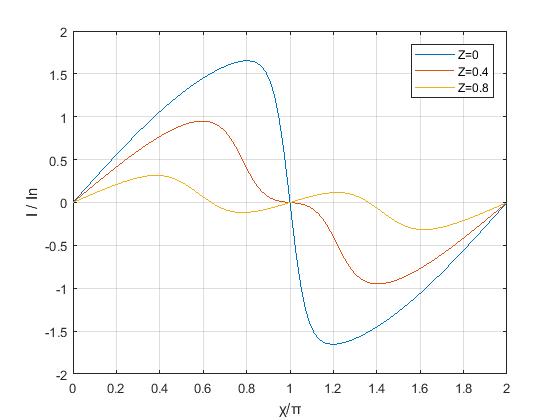}
	
	Figure 3.7: 0-junction to second harmonic due to change of Z  \\(ZZ geometry, $k_{F}d$=10, $\lambda$=0.1) 
\end{center}
\newpage
\par The second step includes the domination of the $\pi$-junction over the second harmonic. In figure (3.8) we present this process. Again, the parameter we change is the scattering interfaces' amplitude:
\bigskip
 
\begin{center}
	\includegraphics[scale=0.5]{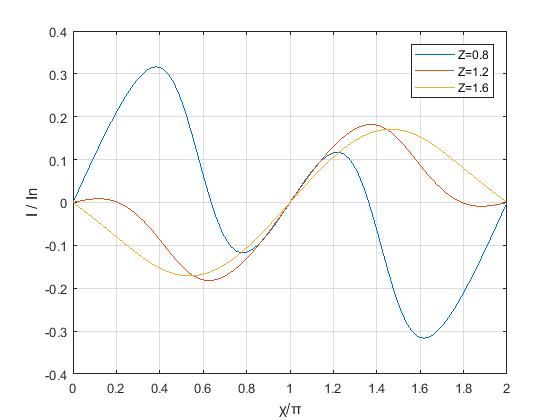}
	
	Figure 3.8: Second harmonic to $\pi$-junction due to change of Z  \\(ZZ geometry, $k_{F}d$=10, $\lambda$=0.1) 
\end{center}

\bigskip
\par So, the two graphs show the 0-$\pi$ transition process in total. As mentioned before, in those graphs we have also applied a normal scattering interface. The transition's existence is a result of the magnetization and is not depended of the normal scattering, but the presence of it makes the curves smoother and the transition easier to observe.
\par  The appearance of the second harmonic can also be achieved by changing the value of $\lambda$ or the 2DEG's length $k_{F}d$. The transition in those cases though is very sensitive to the change of these parameters. A convenient way to watch this process in general is by making a 3-D plot of the phase-current relation for a range of values of either of $\lambda$ or $k_{F}$d. Such plots are shown in the figures below:
\newpage

\begin{center}
	\includegraphics[scale=0.6]{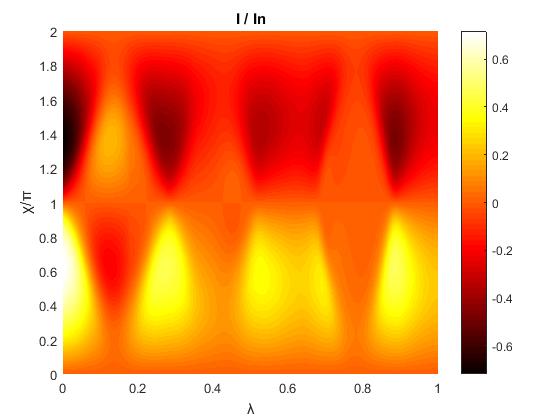}
	
	Figure 3.9: 3-D plot of the Current-Phase Relation for $\lambda$$\in$[0,1] \\(ZZ geometry, $k_{F}d$=10, Z=1)
\end{center}
\bigskip

We can plot separately the Current as a function of the phase for \newline $\lambda\in[0.1,0.2]$ and $\lambda\in[0.4,0.5]$ , which are the fields we have the appearance of the second harmonic. Such plots are shown in figure (3.10):
\bigskip

\begin{center} 
	\includegraphics[scale=0.5]{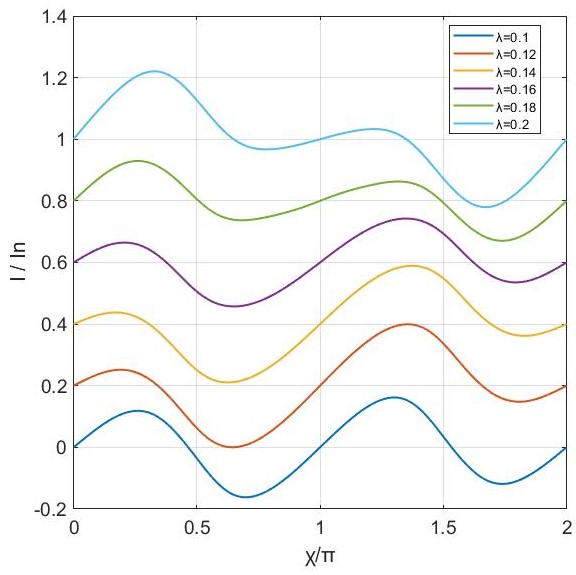}\includegraphics[scale=0.5]{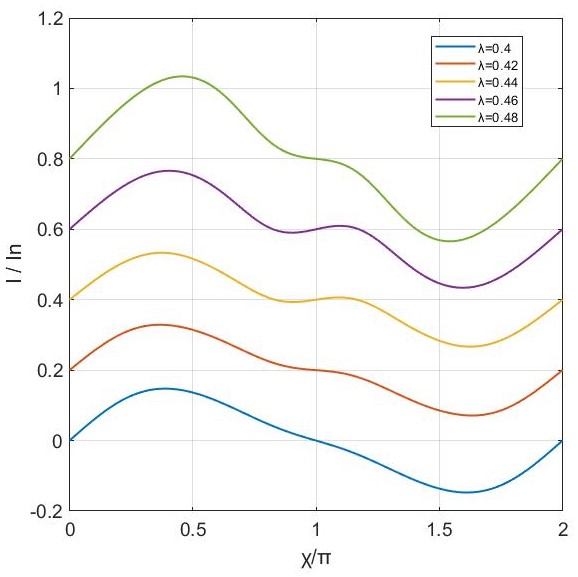}
	
	Figure 3.10: Current-Phase Relation in regions with second harmonic  \\(ZZ geometry, $k_{F}d$=10, Z=1, Displacement=0.2)
\end{center}

\begin{center}
	\includegraphics[scale=0.6]{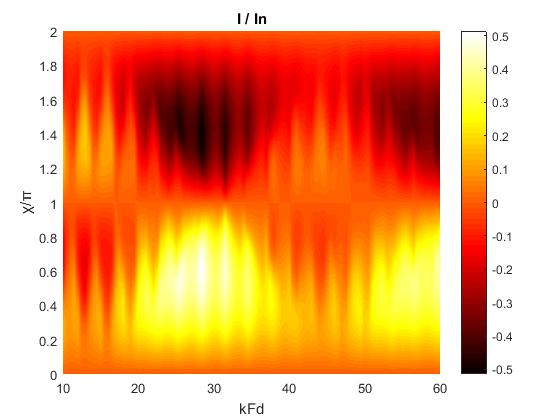}
	
	Figure 3.11: 3-D plot of the Current-Phase Relation for $k_{F}d\in$[10,60] \\(ZZ geometry, $\lambda$=0.1 , Z=1)
\end{center}
\bigskip

Again, we plot separately the Current-phase relation for two of the fields we have the appearance of the second harmonic (figure 3.12). 
\bigskip

\begin{center} 
	\includegraphics[scale=0.5]{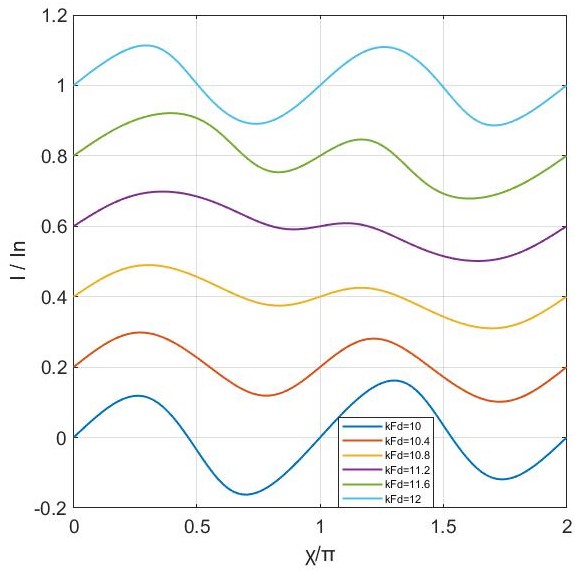}\includegraphics[scale=0.5]{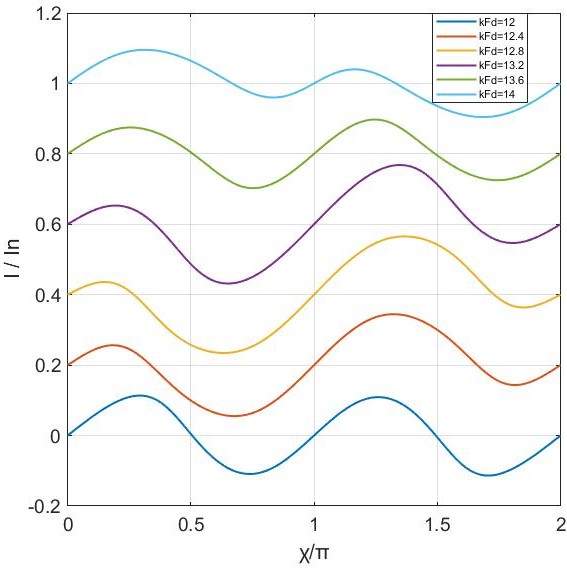}
	
	Figure 3.12: Current-Phase Relation in regions with second harmonic  \\(ZZ geometry, $\lambda$=0.1 , Z=1, Displacement=0.2)
\end{center}
\bigskip
Figure (3.9) represents the current-phase relation over a range of $\lambda$. From this graph we can see that the 0-junction appears periodically as we change the value of $\lambda$. Between two 0-junctions we have the appearance of the second harmonic, which amplitude though decays rapidly for bigger values of $\lambda$ (as seen in figure 3.10). Also, this amplitude is quite weaker than the amplitude of the 0-junction. Another interesting fact, is that we don't see the appearance of $\pi$ junction in this graph.
\par In figure (3.11) we do the same work, for a range of values of $k_{F}$d this time. From the graph's form we see that the dependence from $k_{F}d$ is a more complicated one. If we take a closer look at this graph, we notice that two periods exist in this case. We observe interchange of larger regions (width 10), where the 0-junction is dominant and regions where second harmonic is dominant but interchanges with 0-junction with a small period. We can also notice that, even though the second harmonic appears more frequently , it's amplitude is not as strong as the one of the 0-junction. Thus, when we have the appearance of both of them, the 0-junction is the phenomenon that  dominates the other. In addition, $\pi$-junction doesn't seem to occur in this case either.
\par Finally, in figures (3.13) and (3.14) we show the corresponding graphs for the other three geometries shown in figure (3.6): 

\begin{center} 
	\includegraphics[scale=0.4]{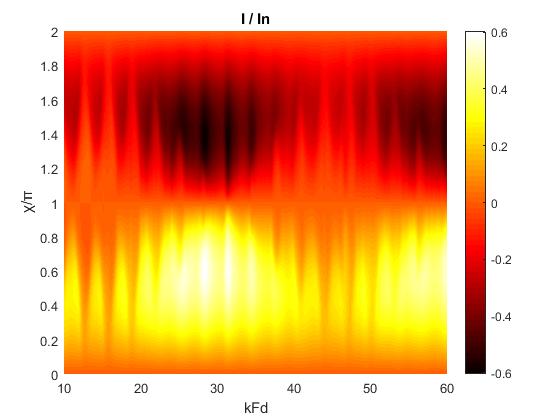}\includegraphics[scale=0.4]{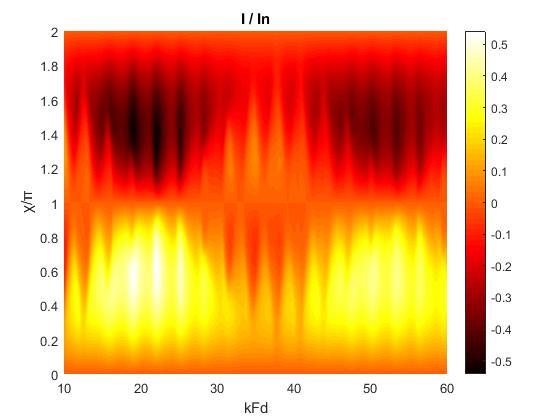}
	\includegraphics[scale=0.4]{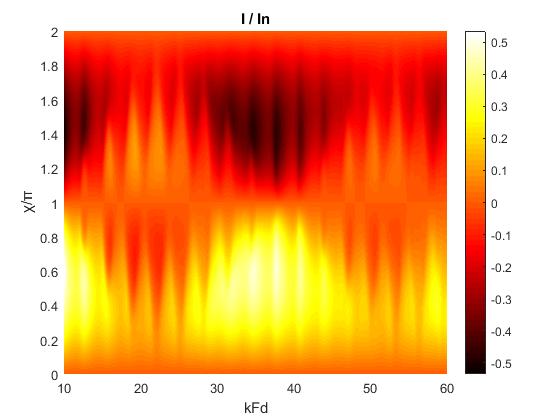}
	
	Figure 3.13: 3-D plot of the Current-Phase Relation for XX, ZX\& XZ Geometries respectively ($k_{F}d\in$[10,60], $\lambda$=0.1 , Z=1)
\end{center}
The general behaviour is quite similar to the case of the ZZ geometry. The graphs are of the same form but for each geometry we have a different, but still small, displacement.  
\par Next we have the graphs for $\lambda$ in figure (3.14):
\begin{center} 
	\includegraphics[scale=0.4]{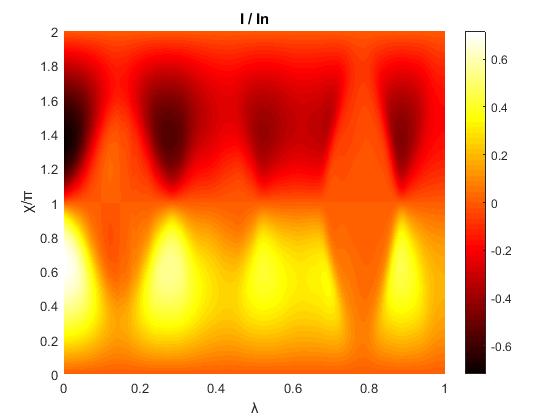}\includegraphics[scale=0.4]{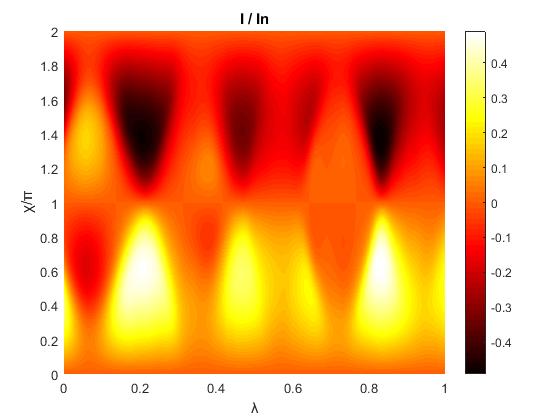}
	\includegraphics[scale=0.4]{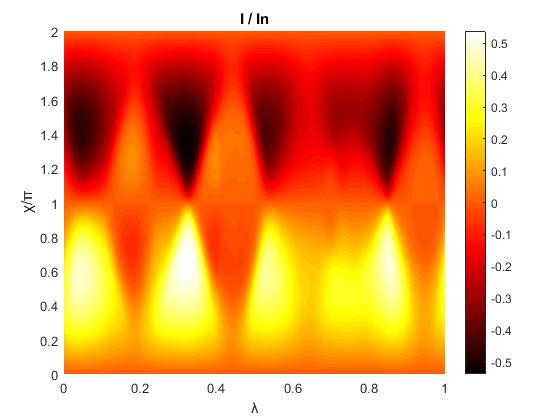}
	
	Figure 3.14: 3-D plot of the Current-Phase Relation for XX, ZX\& XZ Geometries respectively ($\lambda\in$[0,1], $k_{F}d$=10 , Z=1)
\end{center}
\bigskip
\par Again the general behavior is similar to the one in figure (3.9). For different geometries we can observe a displacement in the graph. Due to this displacement, we have the appearance of the 0-junction and the second harmonic in different fields of $\lambda$ for a specific $k_{F}d$ and vice versa.
\bigskip

 \section{The interface magnetization effects II: \\    The Y geometry}

In section 3.2 we studied magnetization geometries which included the Z and X axis and we saw the conditions under which the second harmonic occurs.
\par In this section we study the Y geometry, i.e. junctions with at least one of the magnetization's vectors on the Y axis. We study this geometry separately, because in this case we have the appearance of a variety of new interesting phenomena. The first one is the non-zero Zero-phase current (ZPC), which is the  supercurrent that occurs when the phase difference of the two superconductors is zero: $\chi$=0. Another one is the appearance of a bunch of symmetries, which we examine in the following. 
\par Generally speaking, the current-phase relation is antisymmetric, which is described mathematically as: 
\begin{equation}
I(\chi)=-I(2\pi-\chi)
\end{equation}
\par When we apply, at least one, Y magnetization, though, this (anti)symmetry no longer occurs:
\begin{equation}
I(\chi)\ne -I(2\pi-\chi) \quad\text{(Y magnetization)}
\end{equation}
So, the development of (non-zero) ZPC is a result of the relation (3.2). 
\par Consequently, we shall examine the current-phase relation for these geometries. We shall begin with the geometries that contain one magnetization on the Y axis and the other on Z, considering all possible combinations, in figures (3.15) and (3.16). 
	\begin{figure}
		\centering
	\subfigure[YZ]{\includegraphics[scale=0.6]{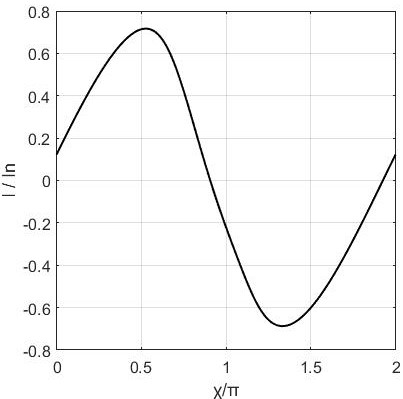}}\subfigure[-YZ]{\includegraphics[scale=0.6]{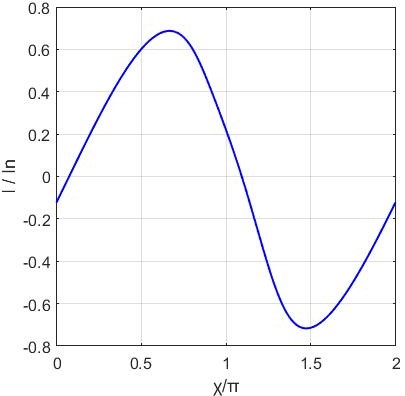}}
	\subfigure[Y-Z]{\includegraphics[scale=0.6]{I_x_GeomYZZ05L05d10}}\subfigure[-Y-Z]{\includegraphics[scale=0.6]{I_x_Geom_YZZ05L05d10}}
	\\
	Figure 3.15: Current-Phase Relation for $\pm$Y$\pm$Z magnetization geometries \\($k_{F}d$=12, $\lambda$=0.6, Z=0.5)
	\end{figure}
	\begin{figure}
		\centering
	\subfigure[ZY]{\includegraphics[scale=0.6]{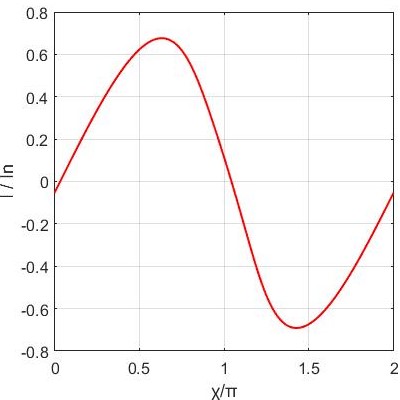}}\subfigure[Z-Y]{\includegraphics[scale=0.6]{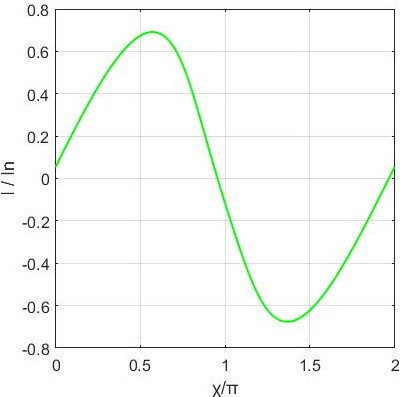}}
	\subfigure[-ZY]{\includegraphics[scale=0.6]{I_x_GeomZYZ05L05d10}}\subfigure[-Z-Y]{\includegraphics[scale=0.6]{I_x_GeomZ_YZ05L05d10}}
	\\
	Figure 3.16: Current-Phase Relation for $\pm$Z$\pm$Y magnetization geometries \\($k_{F}d$=12, $\lambda$=0.6, Z=0.5)
\end{figure}

\par In these figures, one can observe that relation (3.1) no longer applies and also I(0)$\ne$0. For these plots, we have applied a strong spin-orbit coupling constant ($\lambda$=0.6), in order for this "symmetry breaking" to be shown in a more clear way.
\par Another interesting thing in that figure is the fact that each one of the graphs (a)-(h) is related to three of the others with a symmetry operation. The graphs we have given the same color ((a)\&(c), (b)\&(d), etc.) are related with the identical symmetry, namely, the phase-current relation is described from the same function. Each of these graphs occurs form the other, after a 180-degree rotation of the Z magnetization (Z $\rightarrow$ -Z).
\par The other symmetry operation is the center of inversion and relates graphs which occur from a 180-degree rotation of the Y magnetization (Y $\rightarrow$ -Y). For example, we see that, for the current curves in graphs (a) and (b) or (e) and (f), is true that:
\bigskip

\begin{equation}
I_{1}(\chi)=-I_{2}(2\pi-\chi)
\end{equation}
\bigskip

We can also see this relation in figure (3.17) :
\begin{figure}[H]
	\centering
	\subfigure[$\pm$YZ]{\includegraphics[scale=0.4]{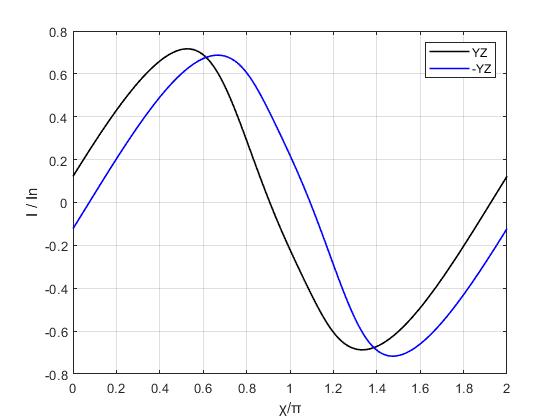}}\subfigure[Z$\pm$Y]{\includegraphics[scale=0.4]{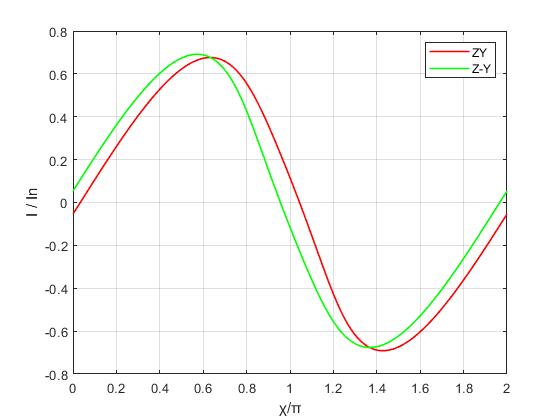}}
\end{figure}
\begin{center}
	Figure 3.17: Symmetry relations of magnetization geometries YZ,-YZ \& ZY,Z-Y  ($k_{F}d$=12, $\lambda$=0.6, Z=0.5)  
\end{center}
\bigskip

\par Next,  we shall examine the case of X and Y magnetization geometries combination, in a same manner. In figures (3.18) and (3.19)  we show the corresponding graphs, which appear to comply with the respective symmetry operations:
\bigskip
\begin{center}
	X $\rightarrow$ -X : Identity, \quad Y $\rightarrow$ -Y : Center of inversion
\end{center}

	\begin{figure}[H]
	\centering
	\subfigure[YX]{\includegraphics[scale=0.6]{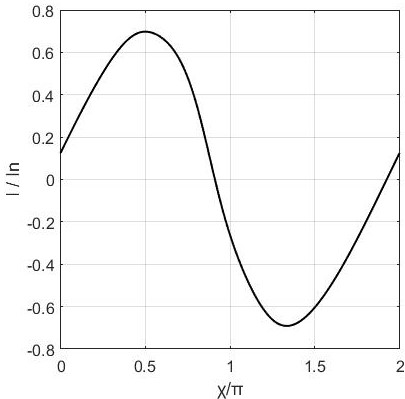}}\subfigure[-YX]{\includegraphics[scale=0.6]{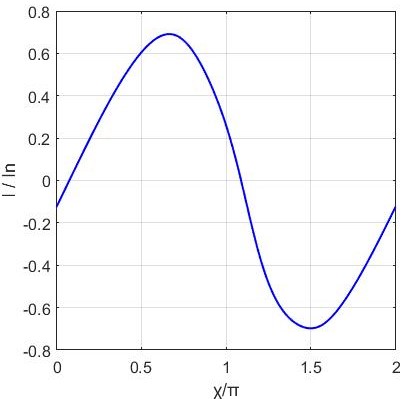}}
	\subfigure[Y-X]{\includegraphics[scale=0.6]{I_x_GeomYXZ05L05d10}}\subfigure[-Y-X]{\includegraphics[scale=0.6]{I_x_Geom_YXZ05L05d10}}
	\\
	Figure 3.18: Current-Phase Relation for $\pm$Y$\pm$X magnetization geometries \\($k_{F}d$=12, $\lambda$=0.6, Z=0.5)
\end{figure}
\begin{figure}[H]
	\centering
	\subfigure[XY]{\includegraphics[scale=0.6]{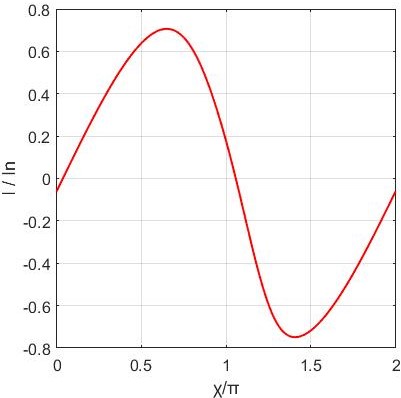}}\subfigure[X-Y]{\includegraphics[scale=0.6]{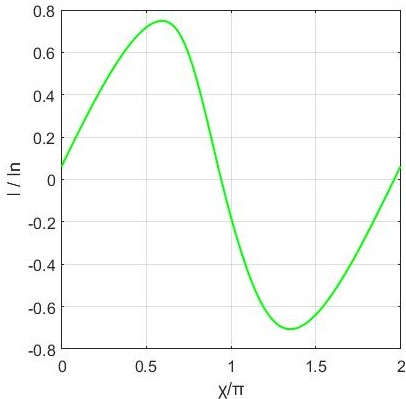}}
	\subfigure[-XY]{\includegraphics[scale=0.6]{I_x_GeomXYZ05L05d10}}\subfigure[-X-Y]{\includegraphics[scale=0.6]{I_x_GeomX_YZ05L05d10}}
	\\
	Figure 3.19: Current-Phase Relation for $\pm$X$\pm$Y magnetization geometries \\($k_{F}d$=12, $\lambda$=0.6, Z=0.5)
\end{figure}

\par Finally, we examine the case in which both of the magnetizations are on the Y axis. We have four combinations this time:
\begin{figure}[H]
	\centering
	\subfigure[YY]{\includegraphics[scale=0.6]{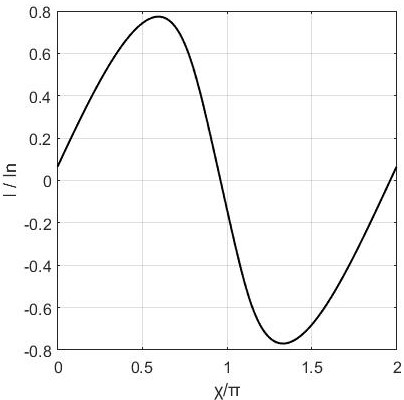}}\subfigure[Y-Y]{\includegraphics[scale=0.6]{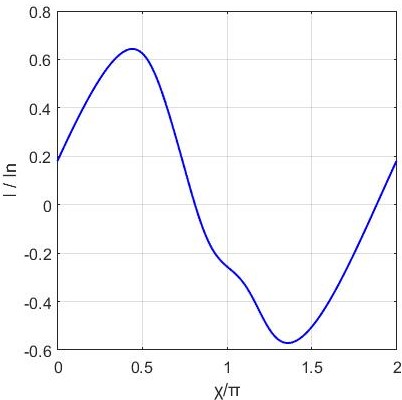}}
	\subfigure[-YY]{\includegraphics[scale=0.6]{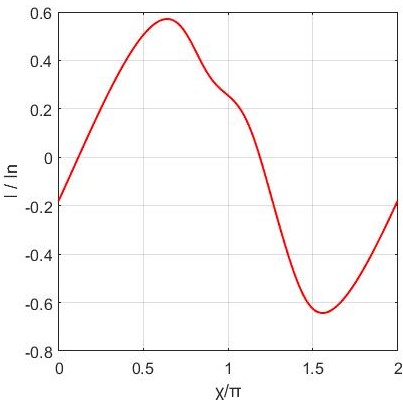}}\subfigure[-Y-Y]{\includegraphics[scale=0.6]{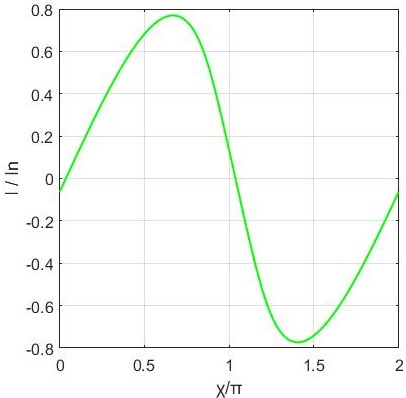}}
	\\
	Figure 3.20: Current-Phase Relation for Y \& Y magnetization geometries \\($k_{F}d$=12, $\lambda$=0.6, Z=0.5)
\end{figure}
\newpage
\par This time, only the symmetry operation of center of inversion is in effect and occurs if we rotate both magnetizations 180 degrees:
\bigskip

\begin{center}
	($\pm$$Y_{1}$ $\rightarrow$ $\mp$$Y_{1}$) \&  ($\pm$$Y_{2}$ $\rightarrow$ $\mp$$Y_{2}$) : Center of inversion
\end{center}
where in the above, the index 1 stands for the first magnetization and 2 for the second. We show the truth of this relation in the graphs below:
\begin{figure}[H]
	\centering
	\subfigure[$\pm$Y$\pm$Y]{\includegraphics[scale=0.4]{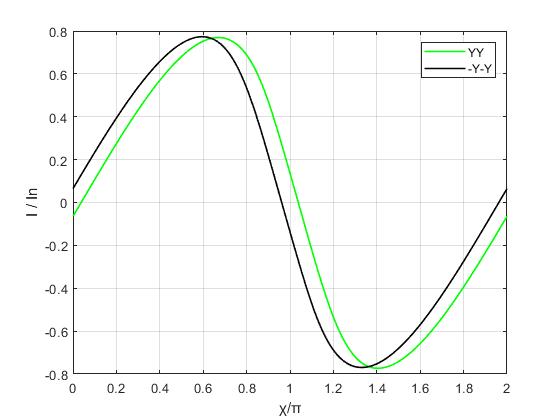}}\subfigure[$\pm$Y$\mp$Y]{\includegraphics[scale=0.4]{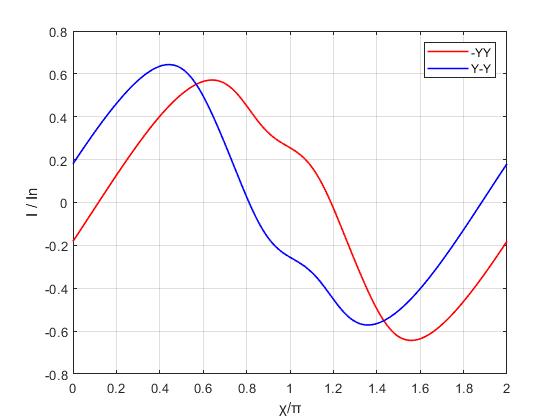}}
\end{figure}
\begin{center}
	Figure 3.21: Symmetry relations of magnetization geometries YY,-Y-Y \& -YY,Y-Y  ($k_{F}d$=12, $\lambda$=0.6, Z=0.5)  
\end{center}
\newpage
\chapter{Properties of Critical Current}
\par In this chapter we study thoroughly the dependence of the Critical (Maximum) supercurrent from the various parameters of our problem, again focusing on the interface magnetization's role. We also examine the contribution to this quantity, as a function of the parallel to the interface momentum $k_{P}$. 
\section{General Study}
One of the most important topics in every electronics problem is the definition of the conditions under which the current is maximized. So, in this first section of the chapter, we make a general study of the way the critical current I$_{c}$ varies according to the value we give to our parameters.
\par Note that for the whole analysis below we consider the absolute value of the critical current. 
\par To begin with, we examine the effect of the (normalized) temperature $\dfrac{T}{T_{c}}$ in figure 4.1.As expected, I$_{c}$ is a decreasing function of T, which tends to 0 when T$\rightarrow$$T_{c}$.

\begin{center}
	\includegraphics[scale=0.5]{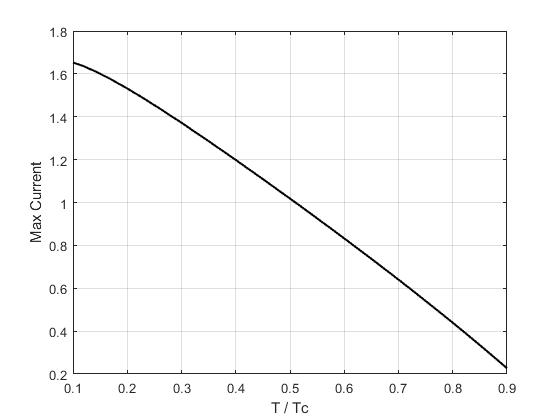}
	
	Figure 4.1: I$_{c}$ as a function of $\frac{T}{T_{c}}$\\ (Z=0, $k_{F}d$=10, $\lambda$=0.1)
\end{center}
\bigskip
\par Next, we shall observe the dependence from the normal scattering interfaces' amplitude Z$_{n}$ (figure 4.2):
\begin{center}
	\includegraphics[scale=0.5]{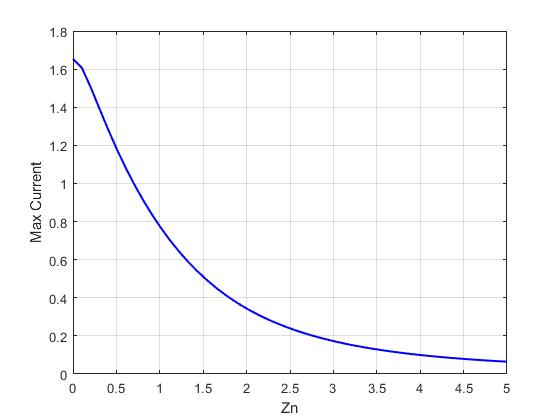}
	
	Figure 4.2: I$_{c}$ as a function of Z$_{n}$\\ (Z$_{m}$=0, $k_{F}d$=10, $\lambda$=0.1)
\end{center}
\par As seen from the graph above, I$_c$ decreases exponentially as we increase Z$_{n}$. 
\par Consequently, we shall examine the relation to the 2DEG's length, $k_{F}d$ (figure 4.3):  
\begin{figure}[H]
\includegraphics[scale=0.45]{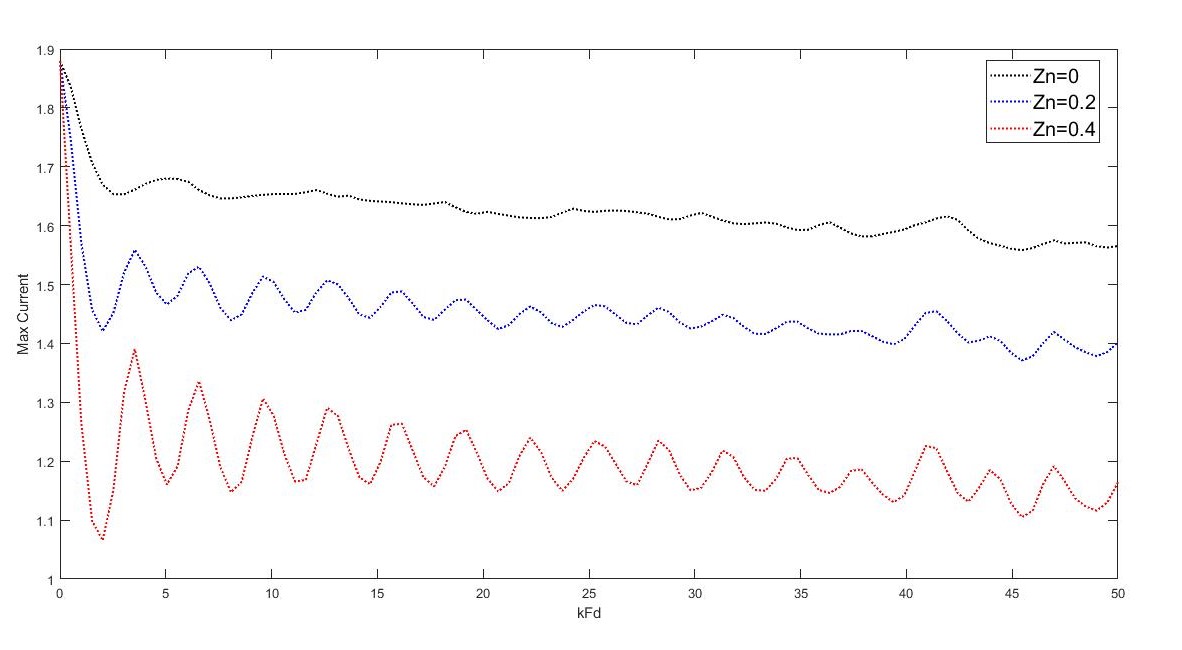}
\end{figure}

\begin{center}	
	Figure 4.3: I$_{c}$ as a function of $k_{F}d$ for various Z$_{n}$\\ (Z$_{m}$=0, $\lambda$=0.1)
\end{center}
\bigskip

In figure (4.3) we see that I$_{c}$ oscillates with a constant period and a, not strictly, decreasing amplitude, as a function of $k_{F}d$. The peaks appear due to normal scattering processes and sharpen for greater values of $Z_{n}$.
\par Finally, we study the dependence from the spin-orbit coupling constant, $\lambda$, in figure (4.4).The curve represents again an oscillation, with two different peaks, which becomes more intense as we increase the value of  $\lambda$. The amplitude is not strictly decreasing, also. Another significant feature is the fact that the three curves, which come from different values of Z$_{n}$, tend to overlap for $\lambda \ge 1.5$.

\begin{center}
	\includegraphics[scale=0.45]{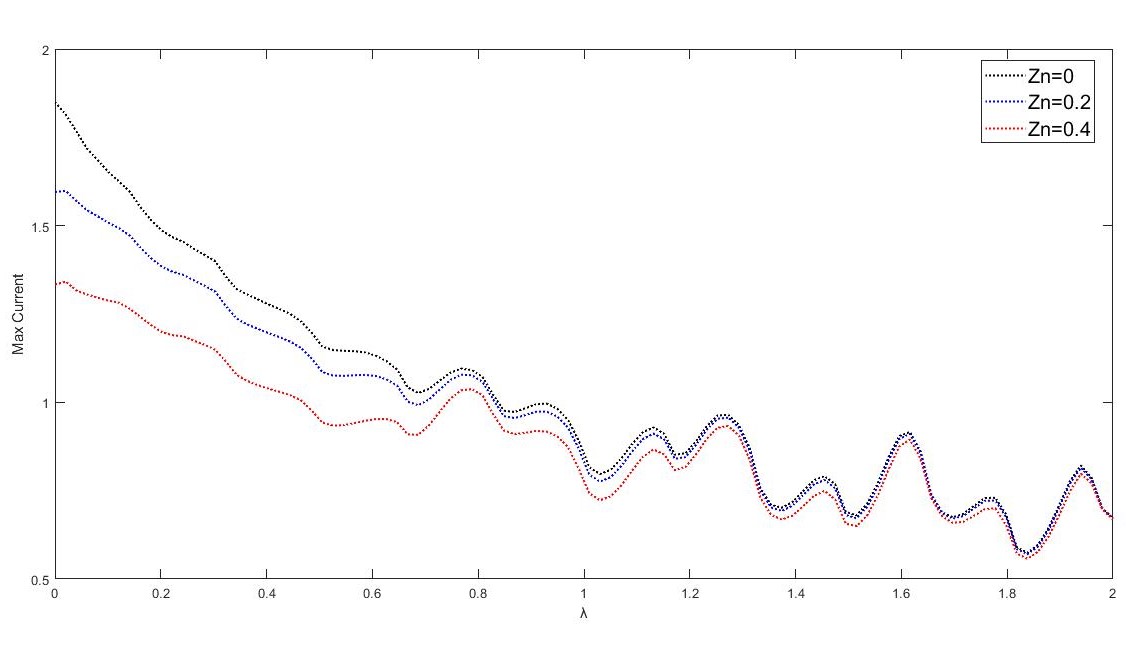}
	
	Figure 4.4: I$_{c}$ as a function of $\lambda$ for various Z$_{n}$\\ (Z$_{m}$=0, $k_{F}d$=10)
\end{center}
\bigskip

\par We can also study the dependence from k$_{F}$d and $\lambda$ from a 3-D plot of ZPC as a function of these parameters. Such plots can be seen in figure (4.5):
\begin{center} 
	\includegraphics[scale=0.40]{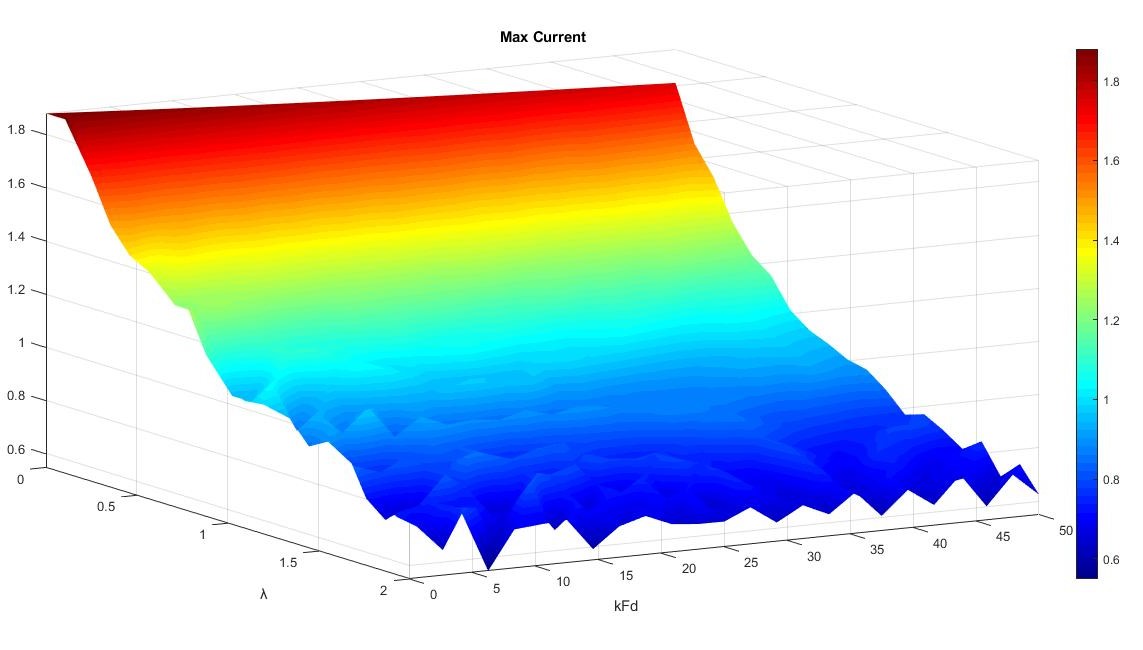}
	
	Figure 4.5: 3-D plot of I$_{c} $ as a function of $k_{F}d$ and $\lambda$  \\(Z=0)
\end{center}

	\section{Max Current as a function of $k_{P}$}
\par In this section of the chapter, we will study the contribution to the Max current of the various parallel wavevectors $k_{P}$, for different values of our parameters. In such diagrams, the integral of the curve, which is the total contribution of all the parallel wavevectors $k_{P}\in[-k_{F},k_{F}]$, gives us the total I$_{c}$:
\begin{equation}
\int_{-k_{F}}^{k_{F}} I(k_{P})dk_{P}=I_{c}
\end{equation}
\ We shall begin our study with showing this contribution graphs for various values of the spin-orbit coupling constant $\lambda$:
\begin{center} 
	\includegraphics[scale=0.45]{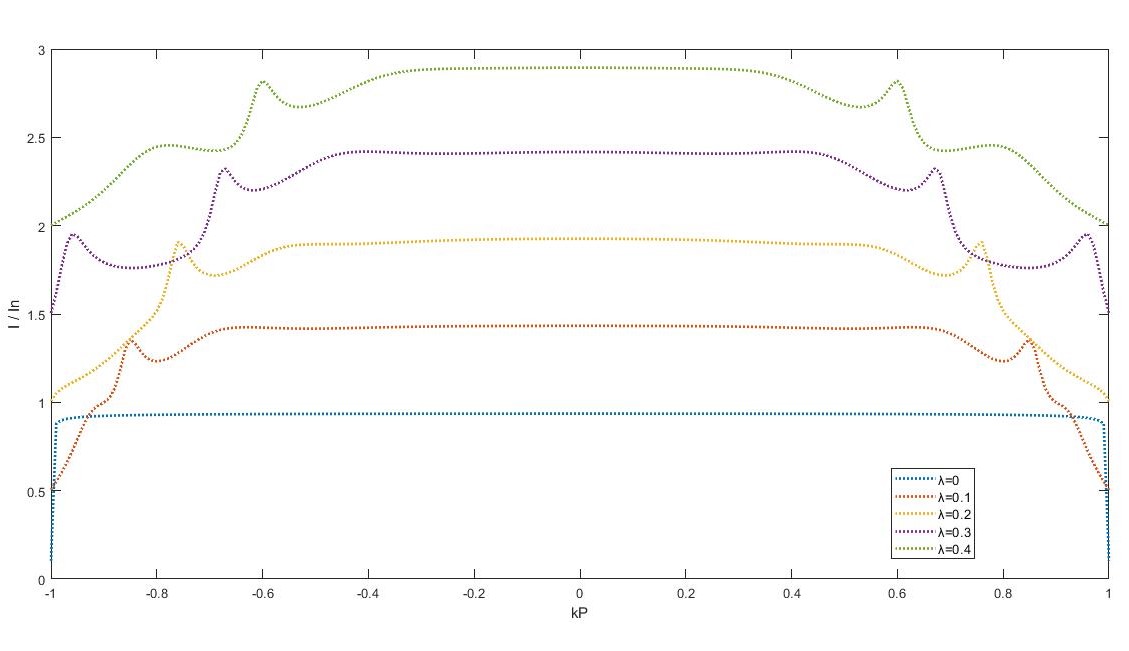}
	
	Figure 4.6: I$_{c}$-$k_{P}$ for various $\lambda$  \\(Z=0, $k_{F}d$=10, Displacement=0.5)
\end{center}
\bigskip
\par For $\lambda$=0 we see that every direction contributes equally to the maximum current, as expected. As we raise the value of $\lambda$, the field of k$_{P}$'s which contribute equally becomes smaller and is always for values near k$_{P}$=0. For greater values of the parallel wavevector, the contribution decays, but there exists no cut-off, as in the case of magnetic layers. We can also observe the presence of an increasing number of peaks, which come from resonances due to normal scattering process. 
\par Next, we may examine how this relation differs when we change the normal scattering interfaces' amplitude, in the presence of a non-zero spin-orbit constant:
\begin{center} 
	\includegraphics[scale=0.45]{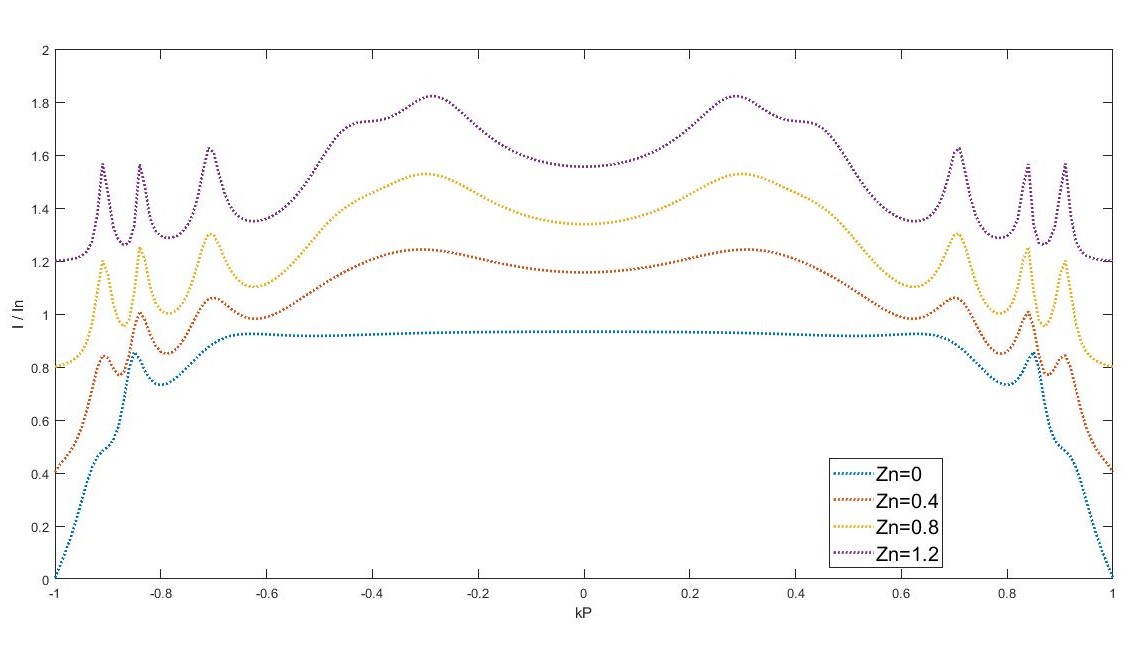}
	
	Figure 4.7: I$_{c}$-$k_{P}$ for various Z$_{n}$  \\(Z$_{m}$=0, $k_{F}d$=10, $\lambda$=0.1, Displacement=0.4)
\end{center}
\bigskip
\par We can notice that when we raise the scattering, the contribution of the incident wave (k$_{P}$=0) becomes lesser. Now the maximum contribution comes from particles with k$_{P} \approx$0.3. We can also observe an increasing number of peaks for greater $\lambda$, which now become sharper due to scattering. Also, for larger Z$_{n}$ these peaks seem to remain the same.
\par Finally, we may study the dependence for the 2DEG's length k$_{F}$d in figure (4.8). We can notice that the change of the length does not change the general behavior of the graph but adds resonant peaks for big values of k$_{P}$.
\begin{center} 
	\includegraphics[scale=0.45]{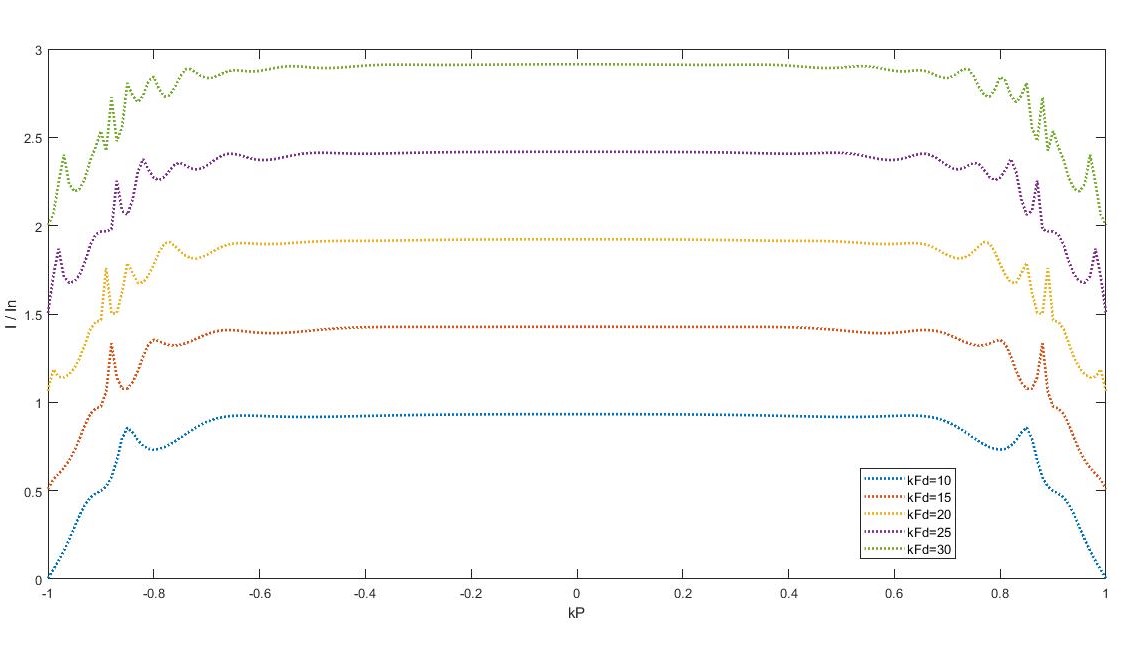}
	
	Figure 4.8: I$_{c}$-$k_{P}$ for various k$_{F}$d  \\(Z=0, $\lambda$=0.1, Displacement=0.5)
\end{center}
\bigskip

\par It is also useful to examine the above dependences from corresponding 3-D plots. Such plots are shown in the figures below: 

\begin{center} 
\includegraphics[scale=0.40]{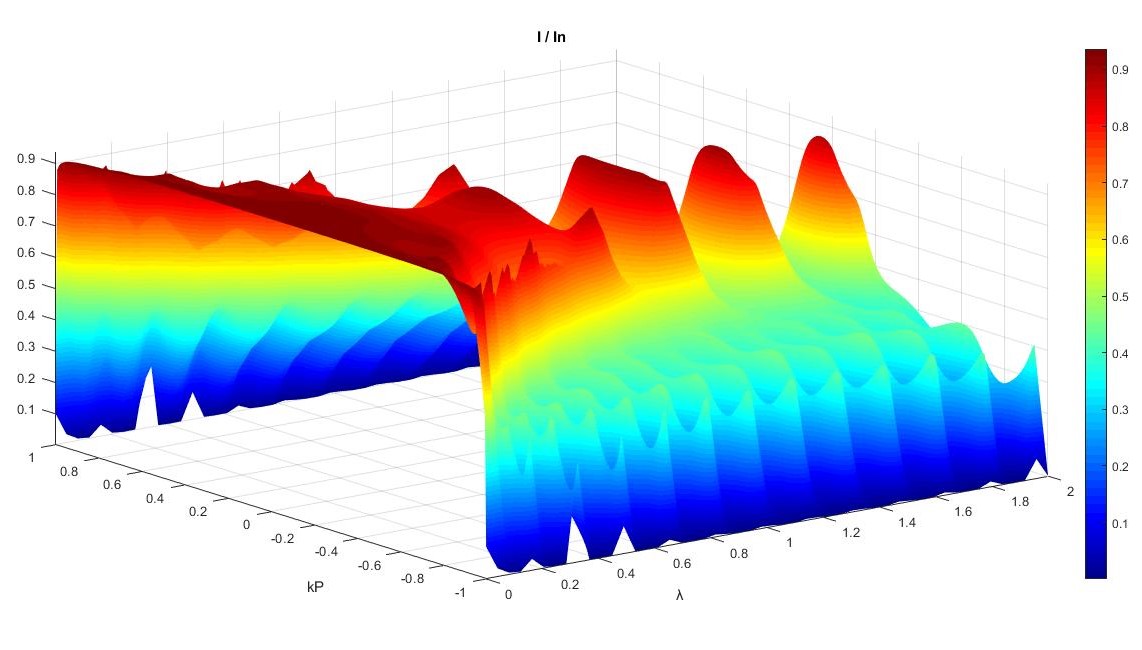}
	
	Figure 4.9: 3-D plot of I$_{c}$ as a function of $k_{P}$ and $\lambda$  \\(Z=0, k$_{F}$d=10)
\end{center}
\bigskip

\begin{center} 
	\includegraphics[scale=0.4]{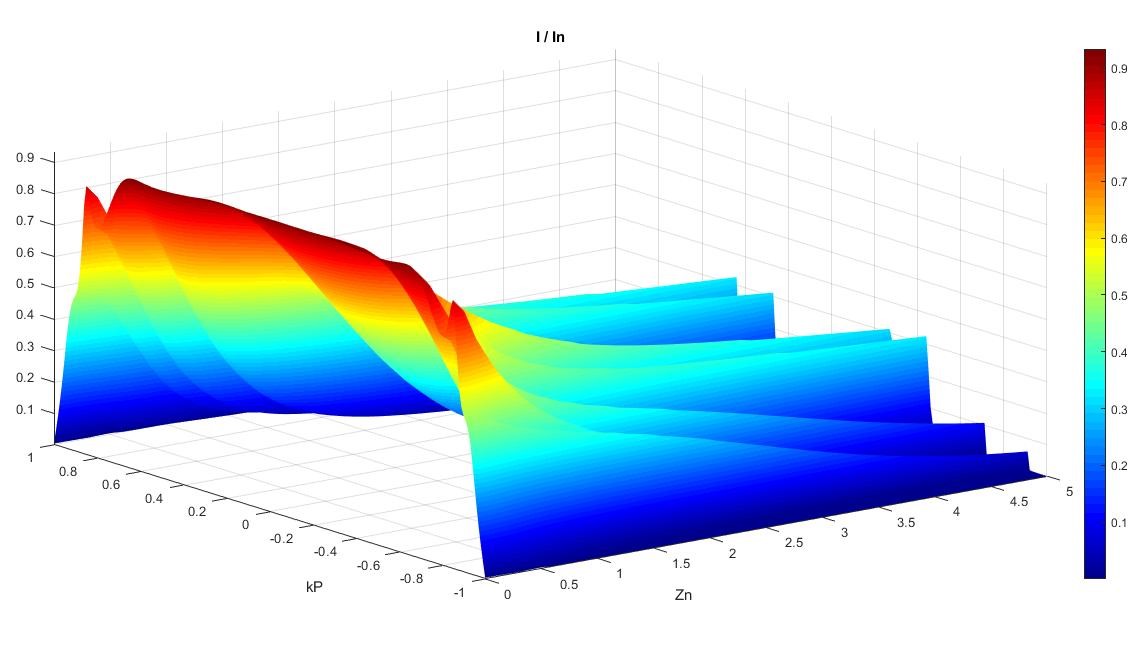}
	
	Figure 4.10: 3-D plot of I$_{c}$ as a function of $k_{P}$ and Z$_{n}$  \\(Z$_{m}$=0, $\lambda$=0.1, k$_{F}$d=10)
\end{center}

\begin{center} 
		\includegraphics[scale=0.4]{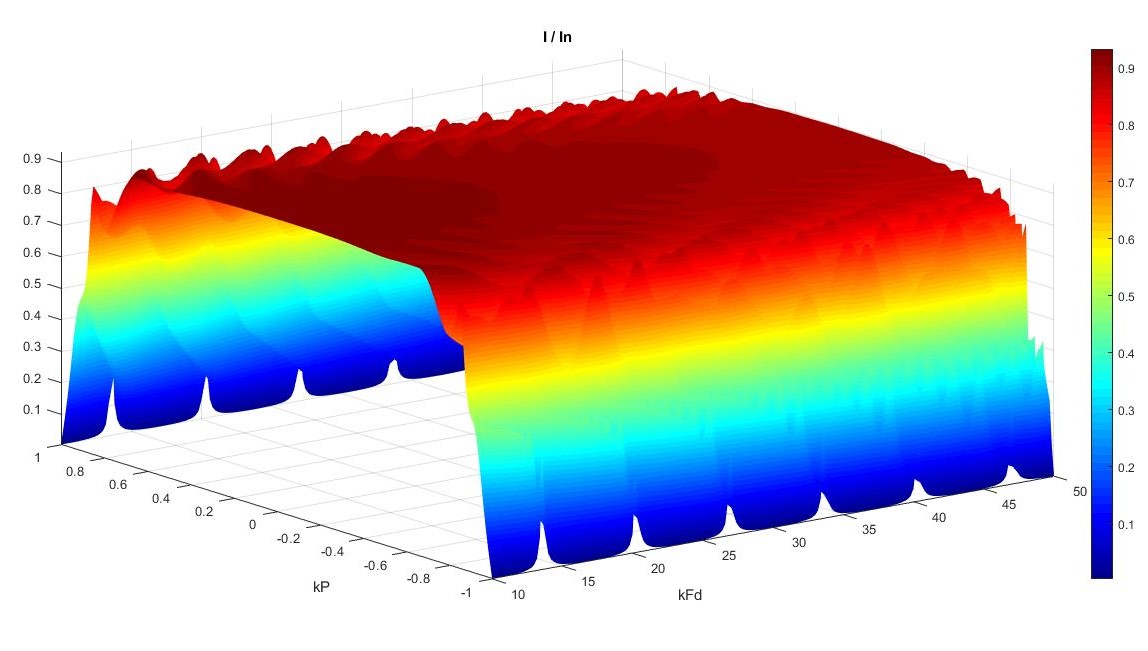}
	
	Figure 4.11: 3-D plot of I$_{c}$ as a function of $k_{P}$ and k$_{F}$d  \\(Z=0, $\lambda$=0.1)
\end{center}

\bigskip

\section{The effect of the magnetizations}
\par In the previous sections we studied how the Maximum current of the junction varies under the change of the most important parameters of our problem. In our analysis we ignored the magnetic (spin-flip) effects of the interfaces, setting Z$_{m}$=0. These effects we will study in this final section of the chapter. The spin-flip scattering in the interfaces gives our problem significant new capabilities, as we also saw in the respective section of chapter 3, without restricting our study in short values of k$_{F}$d, as in the case of magnetic layers.
\par We shall begin our study with showing the dependence of the critical current from the spin-flip scattering strength Z$_{m}$, for our three main magnetization geometries. At first we show this relation for a normal metal junction, for $\lambda$=0, in figure (4.12):
\begin{center} 
	\includegraphics[scale=0.45]{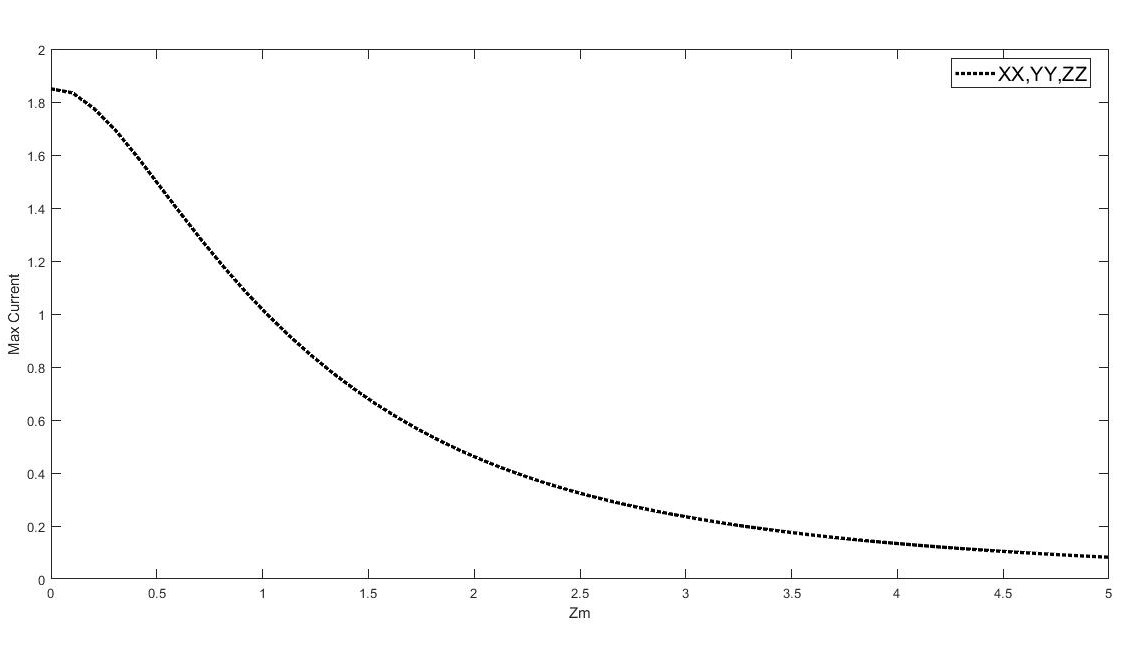}
	
	Figure 4.12: I$_{c}$-Z$_{m}$ for the three main geometries  \\(Z$_{n}$=0, $\lambda$=0, k$_{F}$d=10)
\end{center}
\bigskip

\par From the above figure we can observe that, when $\lambda$=0, the effect of Z$_{m}$ is similar to the effect of Z$_{n}$, shown in figure (4.2): the critical current decays exponentially as we increase the scattering interfaces' strength. Also, another important fact is that the three curves coincide. This is a general feature of the S/N/S junction: the current curve is unchanged when we rotate the whole system by any angle. Thus, only the angle between the two magnetizations, and not their directions, plays role in our study.
\par On the contrary, in an S/2DEG/S junction, i.e. when $\lambda \ne 0$, the system is no longer direction-less. This is a consequence of the Hamiltonian's form in the 2DEG region. As seen from relation (1.2), the inclusion of the triple product now gives direction to our system. So, we expect that when $\lambda \ne 0$ the above three geometries will give us different curves. 
\par In figure (4.13) below we plot the I$_{c}$-Z$_{m}$ relation for those three geometries, for $\lambda$=0.1: 
\bigskip
\begin{center} 
	\includegraphics[scale=0.45]{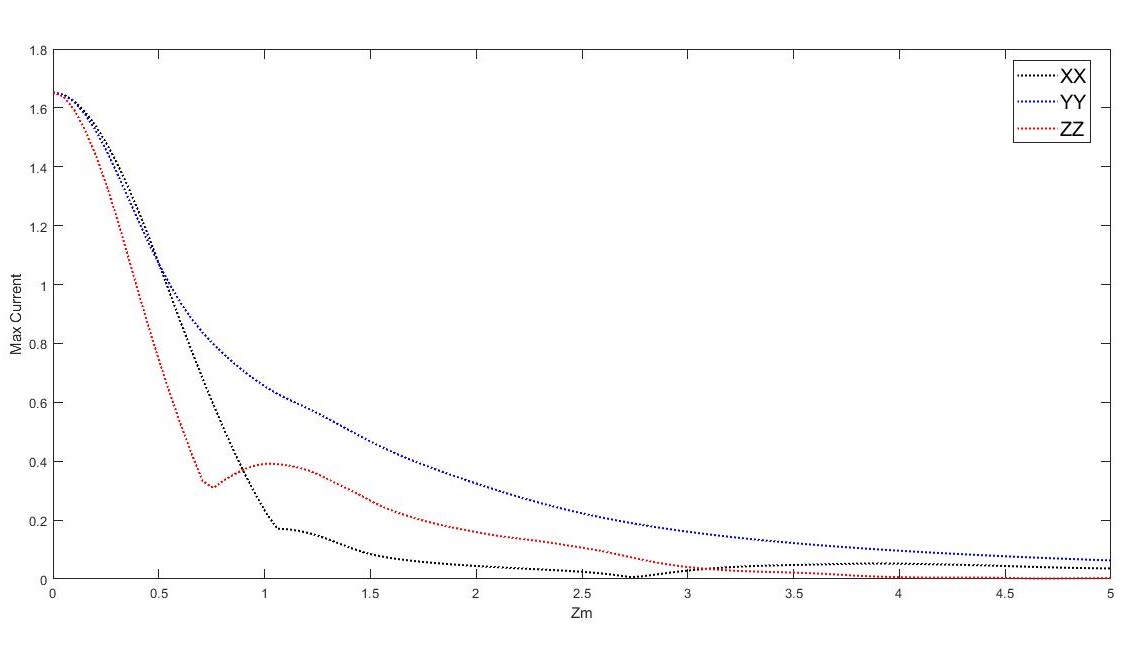}
	
	Figure 4.13: I$_{c}$-Z$_{m}$ for the three main geometries  \\(Z$_{n}$=0, $\lambda$=0.1, k$_{F}$d=10)
\end{center}
We can see that now the three curves show quite different behavior to the change of Z$_{m}$. One can also observe that, at some points, the curves have derivative discontinuities, which result to dips in our graph. In the region of those dips, we have the 0-$\pi$ (or $\pi$-0) transition, we mentioned in the previous chapter. In the figures below, we show the current-phase relation in these regions. Figure (4.14) refers to the peaks in the XX curve.
Next we show the corresponding graph for the ZZ curve peak, in figure (4,15).
\begin{center} 
	\includegraphics[scale=0.45]{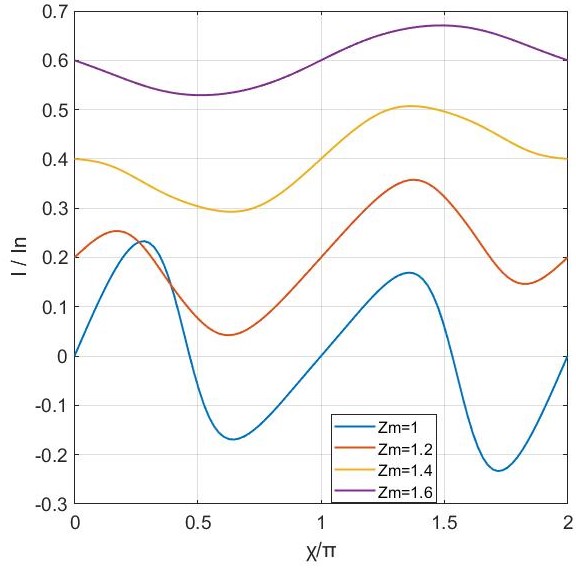}\includegraphics[scale=0.45]{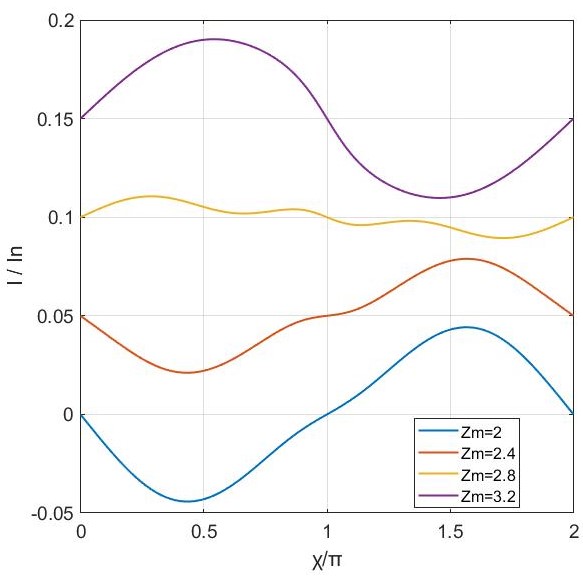}
	
	Figure 4.14: Current-Phase relation for various\\ Z$_{m}$ in [1 , 1.6] and Z$_{m}$ in [2 , 3.2] (transition fields)  \\(XX Geometry, Z$_{n}$=0, $\lambda$=0.1, k$_{F}$d=10, Displacement=0.2 \& 0.05 respectively)
\end{center}
\begin{figure} [H]
	\centering
	\includegraphics[scale=0.45]{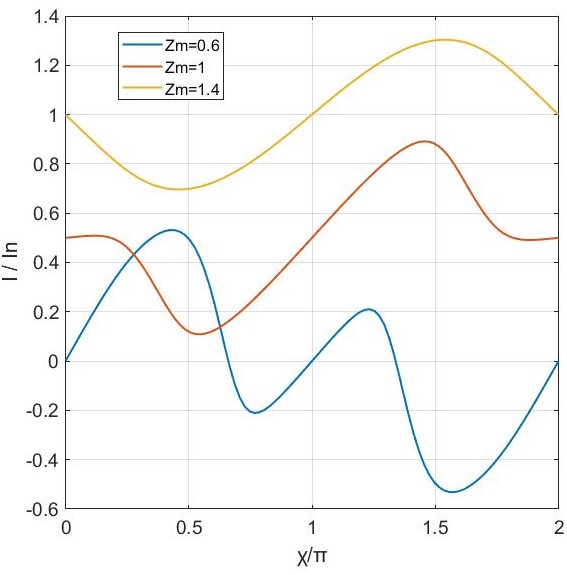}
	
	Figure 4.15: Current-Phase relation for \\various Z$_{m}$ in [0.6 , 1.4]  (transition field)  \\(ZZ Geometry, Z$_{n}$=0, $\lambda$=0.1, k$_{F}$d=10, Displacement=0.5)
\end{figure}
\bigskip
\par These 3 graphs show the process of the transition. The derivative discontinuity appears when the second harmonic becomes stronger and is the harmonic that determines the critical current's amplitude.
\par It is also useful to examine the maximum current that occurs for every magnetization geometry, for specific values of our parameters. In order to do so, we will make 3-D plots of the critical current to the magnetizations' directions. We shall examine the cases in which both our magnetizations are on each of the x-y, x-z and y-z planes and 'rotate' separately one from the other. We use the symbols $\theta_{\nu}$ and $\phi_{\nu}$ for the polar and the azimuthal angle of each magnetization respectively ($\nu=L,R$). So, we have:
\\x-y plane: $\theta_{L}$=$\theta_{R}$=$90^o$ \& $\phi_{L}$,$\phi_{R}$$\in[0,360^o]$
\\x-z plane: ($\phi_{L}$=$\phi_{R}$=$0$ or $\phi_{L}$=$\phi_{R}$=$180^o$)  \& $\theta_{L}$,$\theta_{R}$$\in[0,180^o]$
\\y-z plane: ($\phi_{L}$=$\phi_{R}$=$90^o$ or $\phi_{L}$=$\phi_{R}$=$270^o$)  \& $\theta_{L}$,$\theta_{R}$$\in[0,180^o]$  
\\Following, we show the corresponding plots. Figure (4.16) represents the x-y plane, figure (4.17) the x-z and figure (4.18) the y-z.

\begin{center} 
	\includegraphics[scale=0.5]{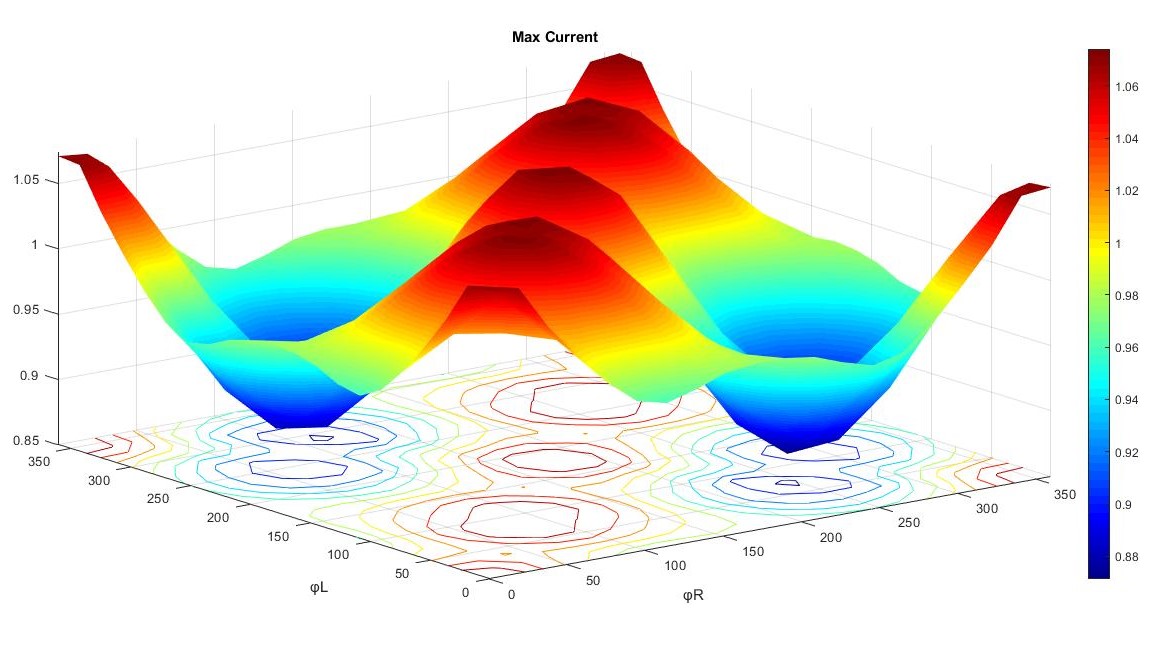}
	
	Figure 4.16: 3-D plot of I$_{c}$ as a function of $\phi_{L}$ and $\phi_{R}$ \\(XY plane, Z$_{m}$=0.5, Z$_{n}$=0, $\lambda$=0.1, k$_{F}$d=10)
\end{center}
As seen from the graph, the critical current maximizes when the vectors become parallel: $\phi_{L}$=$\phi_{R}$ , or anti-parallel: $\phi_{L}$=$\phi_{R} \pm\pi$. The minimum values occur when the two vectors are perpendicular:$\phi_{L}$=$\phi_{R} \pm\dfrac{\pi}{2}$. This is the general behavior, which is interrupted by weaker, but not insignificant, oscillations. We also observe that $I_{c}(\phi_{L},\phi_{R})=I_{c}(\phi_{R},\phi_{L})$.
\par It is also useful to make a Fourier fitting for these graphs, in order to show the explicit dependence from the magnetization angles. So, for figure (4.16) we have:
\begin{equation}
\centering
I_{c}\approx 0.988+0.068cos(\phi_{L}-\phi_{R})-0.023cos(\phi_{L}+\phi_{R})+0.021cos(2(\phi_{L}+\phi_{R}))
\end{equation}
\bigskip

\par Next, we have the plots for the x-z plane in figure (4.17). In this figure, (4.17.a) diagram corresponds to $\phi_{\nu}$=0, while (4.17.b) to $\phi_{\nu}$=180$^o$. This time, the maximum values occur when $\theta_{L}=\theta_{R} \pm\pi$, and the minimum for $\theta_{L}=\theta_{R}$. In addition, we have the symmetry relation: \\$I_{c}(\theta_{L},\theta_{R},\phi_{\nu}=0)=I_{c}(\theta_{R},\theta_{L},\phi_{\nu}=\pi)$.
\begin{center} 
	\includegraphics[scale=0.5]{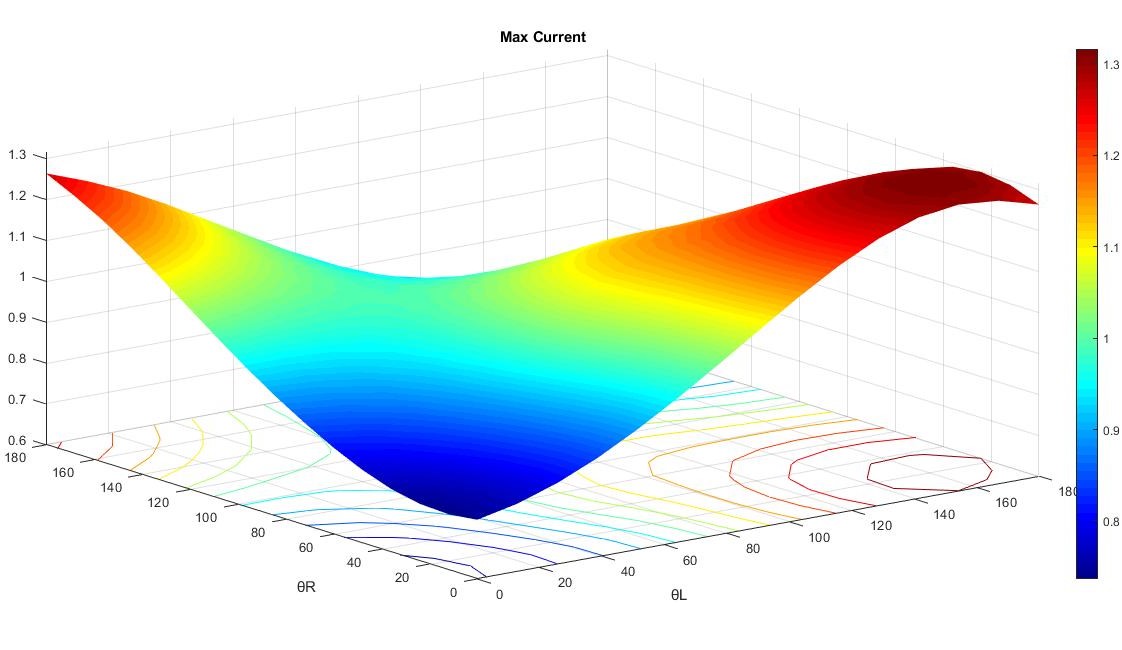}
	
	(4.17.a):$\phi_{\nu}$=0
	
	\includegraphics[scale=0.5]{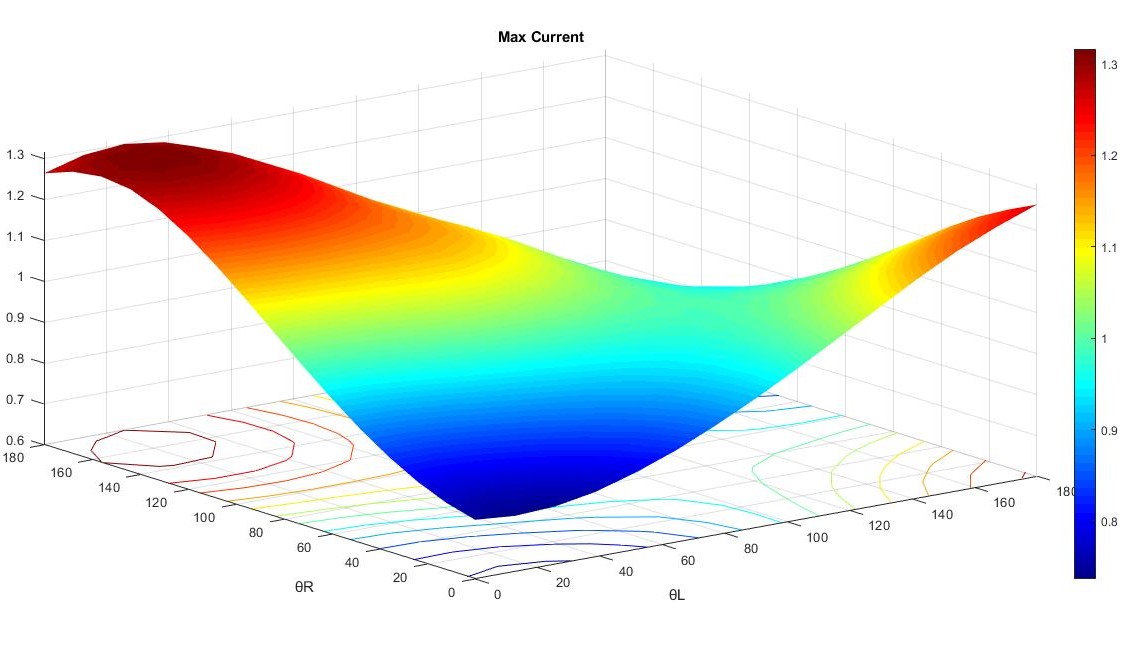}
	
	(4.17.b):$\phi_{\nu}$=180$^o$
	\bigskip
	
	Figure 4.17: 3-D plots of I$_{c}$ as a function of $\theta_{L}$ and $\theta_{R}$ \\(XZ plane, Z$_{m}$=0.5, Z$_{n}$=0, $\lambda$=0.1, k$_{F}$d=10)
\end{center}
\par The Fourier fittings for these graphs are:
\begin{equation}
a: I_{c}\approx 1-0.097cos(\theta_{L}-\theta_{R})+0.1sin(\theta_{L}-\theta_{R})-0.154cos(\theta_{L}+\theta_{R})
\end{equation}
\begin{equation}
b: I_{c}\approx 1-0.097cos(\theta_{L}-\theta_{R})-0.1sin(\theta_{L}-\theta_{R})-0.154cos(\theta_{L}+\theta_{R})
\end{equation}
\par Finally, in figure (4.18) we show the graphs for the y-z plane. Graph (4.18.a) correspond to $\phi_{\nu}=90^o$ and (4.18.b) correspond to $\phi_{\nu}=270^o$. Again, the maximum values appear when $\theta_{L}=\theta_{R} \pm\pi$, and the minimum for $\theta_{L}=\theta_{R}$. 
Also, the two graphs are symmetric to the change $\theta_{L}\leftrightarrow\theta_{R}$ and identical: $I_{c}(\theta_{L},\theta_{R})=I_{c}(\theta_{R},\theta_{L}),\phi_{\nu}=\dfrac{\pi}{2},\dfrac{3\pi}{2}$  \\$I_{c}(\theta_{L},\theta_{R},\phi_{\nu}=\dfrac{\pi}{2})=I_{c}(\theta_{L},\theta_{R},\phi_{\nu}=\dfrac{3\pi}{2})$.
\begin{center} 
	\includegraphics[scale=0.45]{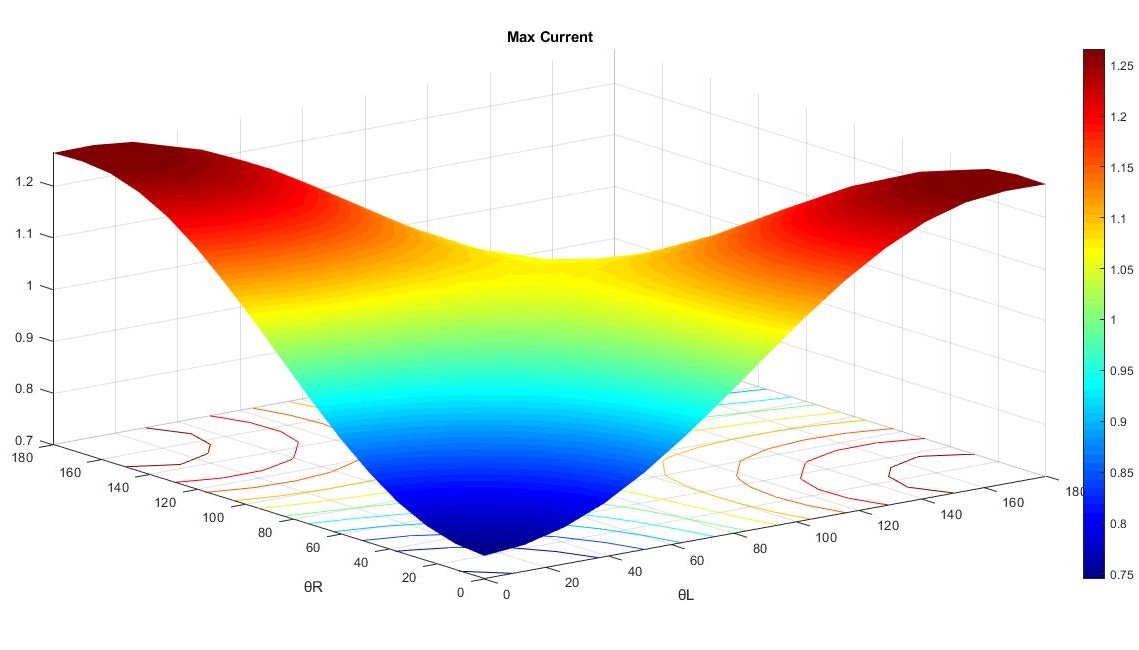}
	
	(4.18.a):$\phi_{\nu}$=90$^o$
	
	\includegraphics[scale=0.45]{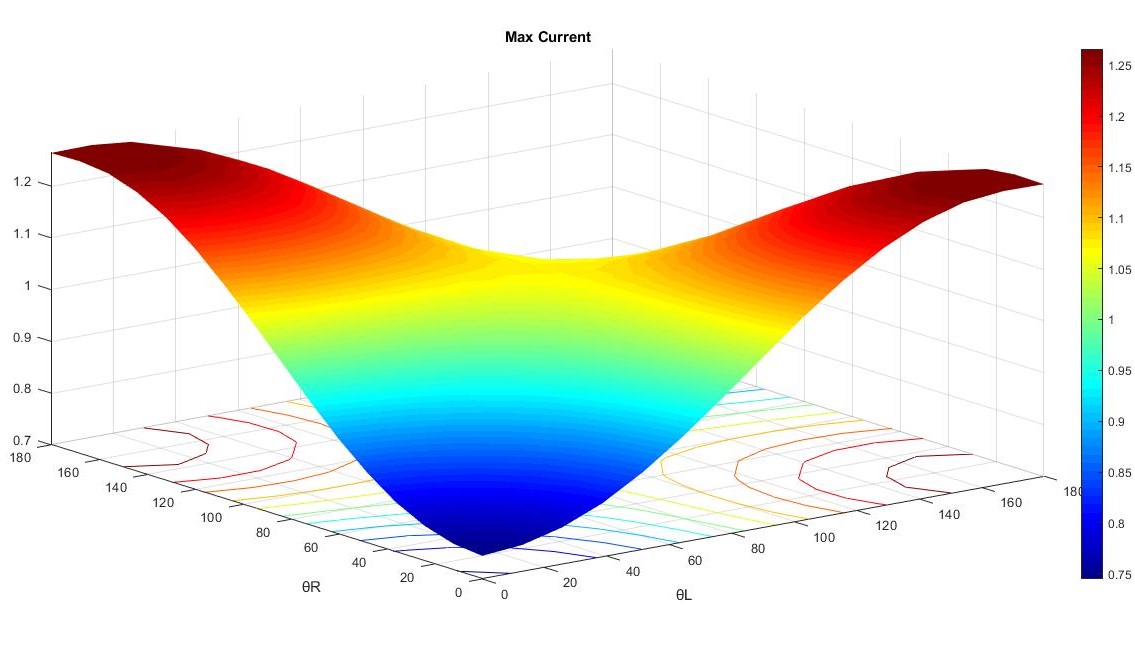}
	
	(4.18.b):$\phi_{\nu}$=270$^o$
	
	\bigskip
	
	Figure 4.18: 3-D plots of I$_{c}$ as a function of $\theta_{L}$ and $\theta_{R}$\\(YZ plane, Z$_{m}$=0.5, Z$_{n}$=0, $\lambda$=0.1, k$_{F}$d=10)
\end{center}
\par The Fourier fitting for both of these graphs is:
\begin{equation}
 I_{c}\approx 1.022-0.109cos(\theta_{L}-\theta_{R})-0.167cos(\theta_{L}+\theta_{R})-0.004cos(2(\theta_{L}+\theta_{R}))
\end{equation}
\par We must note that, in all the above fittings (relations (4.2)-(4.5)), only the terms which are a function of the difference of the corresponding angles ($f(\phi_{L}-\phi_{R})$ or $f(\theta_{L}-\theta_{R})$), as well as the constant terms, appear when we set $\lambda=0$. The other terms (which are of the form $g(\phi_{L}+\phi_{R})$ or $g(\theta_{L}+\theta_{R})$) appear due to the 2DEG, which as we mentioned before makes the system direction-dependent.
\chapter{Properties of the Zero-phase Current}
\par In this chapter we study thoroughly the dependence of the Zero-phase supercurrent (ZPC) from the various parameters of our problem, again focusing on the interface magnetization's role. We also examine the contribution to this quantity, as a function of the parallel to the interface momentum $k_{P}$. 

\section{General Study}
As mentioned before, the ZPC is the  supercurrent that occurs when the phase difference of the two superconductors is zero: $\chi$=0. Thus, ZPC=I(0). In the previous chapter, we saw that ZPC $\ne$ 0, in the case we apply at least one magnetization on the Y axis. So, next we shall study how this ZPC is related with our parameters, in the case of a YY geometry.
\par To begin with, we examine the effect of the (normalized) temperature $\dfrac{T}{T_{c}}$ in figure 5.1. As expected, the ZPC is a decreasing function of T, which tends to 0 when T$\rightarrow$$T_{c}$. Of course, the Critical current vanishes as well in these temperatures.
\begin{figure}[H]
	\centering
	\includegraphics[scale=0.7]{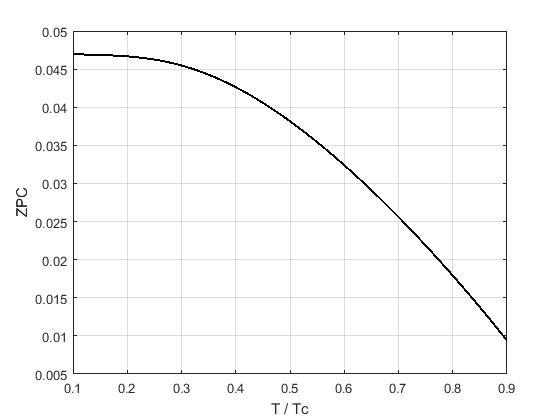}
	
	Figure 5.1: ZPC as a function of $\frac{T}{T_{c}}$\\ (YY geometry, $k_{F}d$=10, $\lambda$=0.5, Z=0.5)
\end{figure}
\bigskip
\par Next, we shall observe the dependence from the scattering interfaces' amplitude (figure 5.2). We set $Z_{n}=Z_{m}=Z$ for this graph. We can see that the ZPC increases steadily, until it reaches a critical value of Z ($Z_{c}$$\approx$0.8), and then, it appears to decay exponentially for greater values of Z. This form has a physical explanation: we have seen that the appearance of the ZPC is an effect of the interface spin-flip scattering. For Z=0, I(0)=0. So, when we increase Z from 0 to greater values, the ZPC increases. When we apply a strong scattering interface though, the damping effect reduces the current's amplitude. So, we expect that for Z$>>$1, I$\rightarrow$0, in general. Thus, the ZPC will have a maximum for an interim value of Z.
\begin{figure}[H]
	\centering
	\includegraphics[scale=0.7]{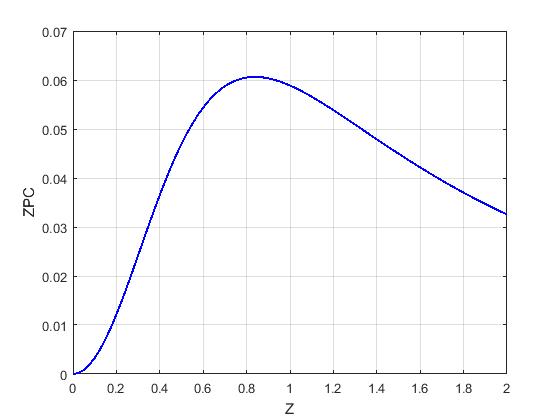}
	
	Figure 5.2: ZPC as a function of Z\\ (YY geometry, $k_{F}d$=10, $\lambda$=0.5)
\end{figure}
\bigskip
\par Following, we shall study the relation to the 2DEG's length, $k_{F}d$ (figure 5.3). In this figure we see that the ZPC oscillates with a constant period and not strictly decreasing amplitude, as a function of $k_{F}d$. The maximum peaks are sharp for small values of the length, while for greater values they become widened and a second peak weaker peak appears next to the maximum.
\begin{figure}[H]
	\centering
	\includegraphics[scale=0.7]{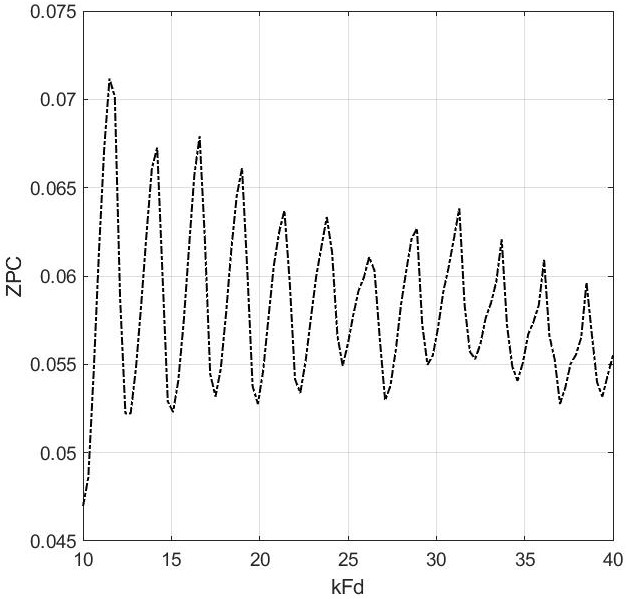}
	
	Figure 5.3: ZPC as a function of $k_{F}d$\\ (YY geometry, $\lambda$=0.5, Z=0.5)
\end{figure}
\bigskip
\par Finally, we examine the dependence from the spin-orbit coupling constant, $\lambda$, in figure (5.4). The curve represents again an oscillation, which amplitude now, though, increases for small values of $\lambda$ and then, for $\lambda$ $\geqslant$ $\lambda_{c}$ ($\lambda_{c}\approx0.62$) it becomes a decreasing function of $k_{F}d$. Similar to the case of Z (figure 5.2), we expect such a general behavior, as we know that I(0)=0 for $\lambda$=0 and $\lambda$$>>$1 (due to mismatch), and also I(0)$\ne$0, for $\lambda$ $\in$(0,1).
\begin{figure}[H]
	\centering
	\includegraphics[scale=0.7]{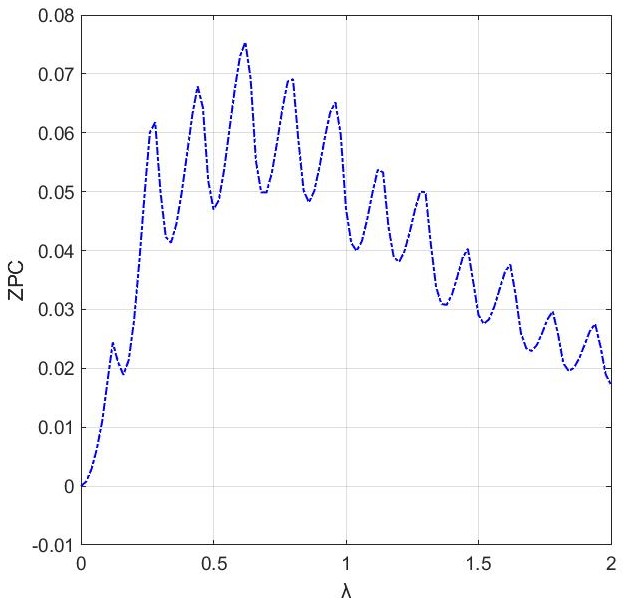}
	
	Figure 5.4: ZPC as a function of $\lambda$\\ (YY geometry, $k_{F}d$=10, Z=0.5)
\end{figure}

\bigskip

\section{ZPC as a function of $k_{P}$}	

\par In this second part of the chapter, we will study the contribution to ZPC of the various parallel wavevectors $k_{P}$, for different values of our parameters. In such diagrams, the integral of the curve, which is the total contribution of all the parallel wavevectors $k_{P}\in[-k_{F},k_{F}]$, gives us the total ZPC:
\begin{equation}
\int_{-k_{F}}^{k_{F}} I(k_{P})dk_{P}=I(\chi=0)
\end{equation}
\ We shall begin our study with showing this contribution graphs for various scattering interfaces' amplitude ($Z=Z_{n}=Z_{m}$), again for YY geometry:
\begin{center} 
	\includegraphics[scale=0.45]{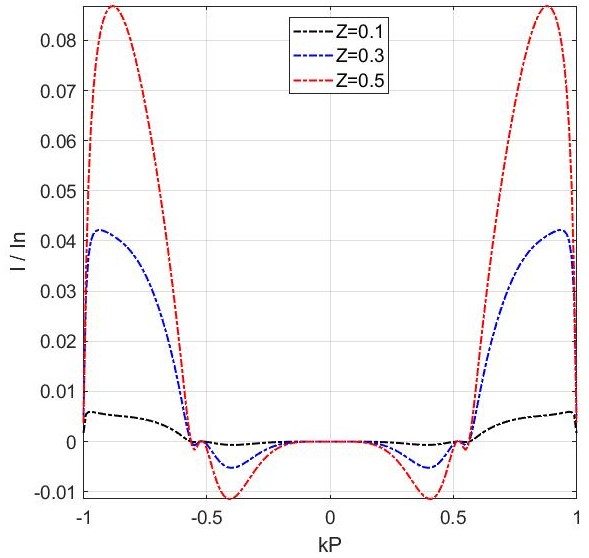}\includegraphics[scale=0.45]{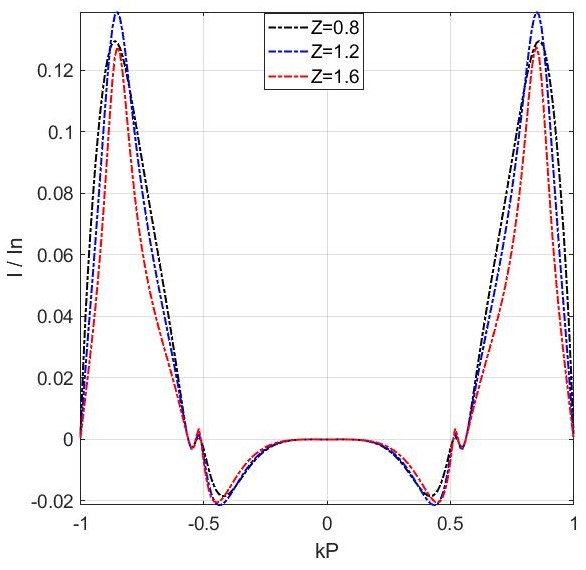}
	
	Figure 5.5: ZPC-$k_{P}$ for various Z  \\(YY geometry, $\lambda$=0.5, $k_{F}d$=10)
\end{center}
\bigskip
\par We see that all the curves have a similar pattern: the contribution becomes important only for $k_{P} > 0.5$, where we have two strong peaks. The peaks become sharper when we increase the scattering amplitude, as expected, and also their amplitude follows the relation shown in figure (5.2). For values of $k_{P}$ near 0, we have no contribution, which means that a particle with perpendicular momentum cannot contribute to the appearance of the ZPC. We can also observe two other weaker and negative peaks for $k_{P}\approx\pm0.4$. This means that particles with that values of $k_{P}$ backscatter and thus, they lower the value of I(0).
\par Next, we may observe how this relation differs when we change the spin-orbit coupling constant $\lambda$:
\begin{center} 
	\includegraphics[scale=0.45]{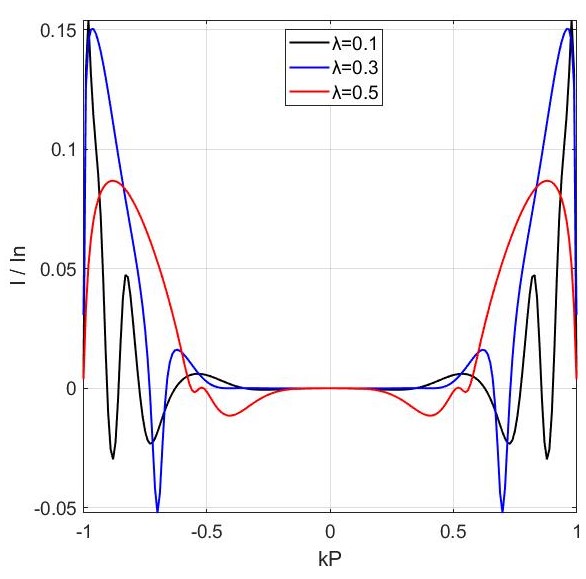}\includegraphics[scale=0.45]{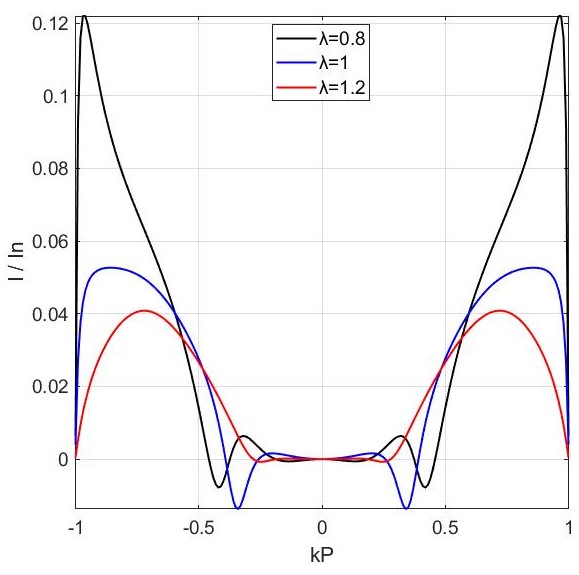}
	
	Figure 5.6: ZPC-$k_{P}$ for various $\lambda$  \\(YY geometry, Z=0.5, $k_{F}d$=10)
\end{center}
\bigskip
\par In this graphs, the values of $k_{P}$ we saw that contribute in figure (5.5), i.e. the peaks, now seem to broaden and also oscillate in amplitude with $\lambda$. We examine this periodical behavior in figure (5.7) below:
\begin{center} 
	\includegraphics[scale=0.5]{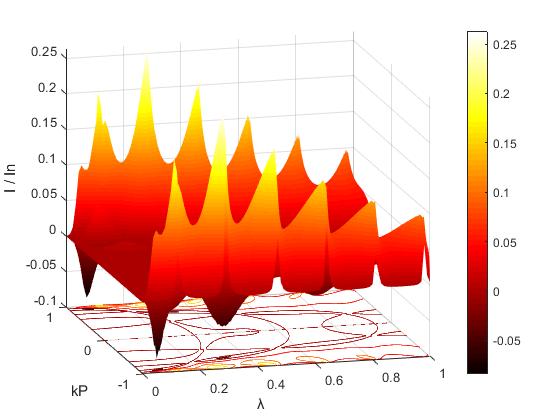}
	
	Figure 5.7: 3-D plots of ZPC-$k_{P}$ for $\lambda \in [0,1]$  \\(YY geometry, Z=0.5, $k_{F}d$=10)
\end{center}
From this graphs we can see that the mentioned behaviour of the various $k_{P}s$ is a periodical function of $\lambda$. The most significant feature is that the peaks' amplitude, which appear near $k_{P} \approx \pm0.85$, seems to maximize periodically, with a constant period $\Delta\lambda \approx 0.17$. These peaks are the same with the ones we saw in figure (5.4).  We can also observe that, the range of the contributing $k_{P}$ is an increasing function of $\lambda$, which tends to become constant for greater values of $\lambda$.
\par Next, we may examine the dependence from $k_{F}d$ in a similar way:
\begin{center} 
	\includegraphics[scale=0.7]{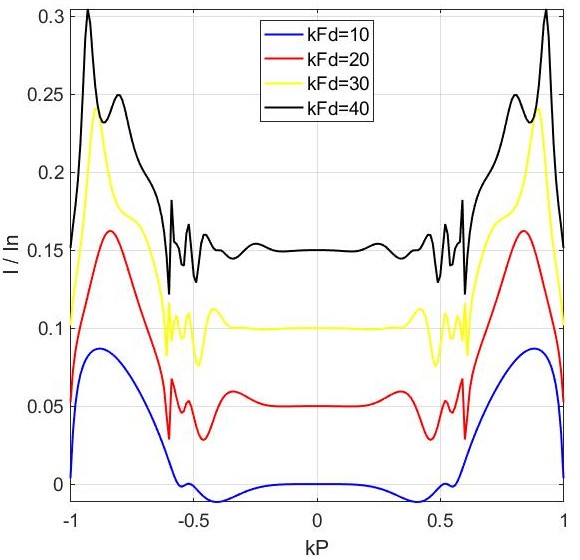}
	
	Figure 5.8: ZPC-$k_{P}$ for various $k_{F}d$  \\(YY geometry, Z=0.5, $\lambda$=0.5, Displacement=0.05)
\end{center}
\newpage

Following we have the 3-D graph: 

\begin{center} 
	\includegraphics[scale=0.6]{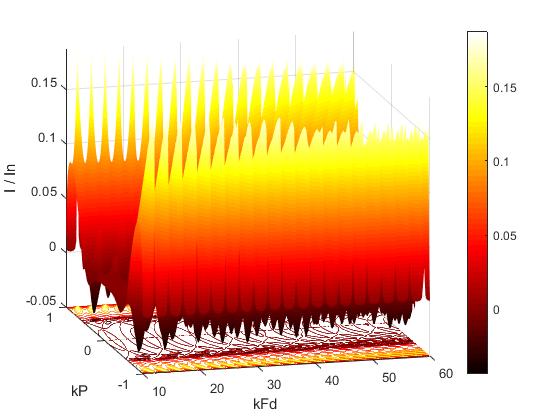}
	
	Figure 5.9: 3-D plots of ZPC-$k_{P}$ for $k_{F}d \in [10,60]$  \\(YY geometry, Z=0.5, $\lambda$=0.5)
\end{center}
Again the dependence is periodical, and the form is quite more complicated than before.
\par Finally, we shall study the corresponding graphs for the various magnetization symmetries that we studied in section 3.3. In figure 5.11 below, we plot the ZPC-k$_{P}$ relation for symmetric geometries.As expected, when the two Current-phase graphs are related with the Center of inversion symmetry, the respective ZPC-k$_{P}$ graphs are antisymmetric.
\newpage

\begin{center} 
	\includegraphics[scale=0.4]{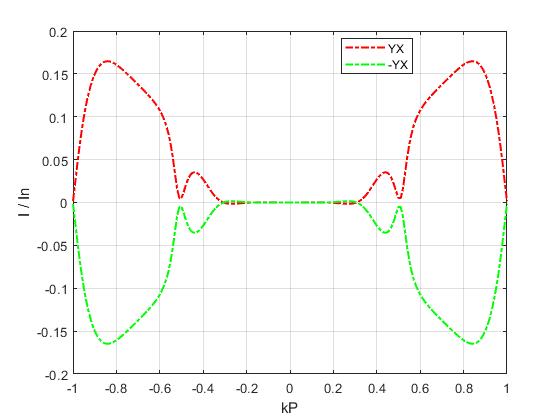}\includegraphics[scale=0.4]{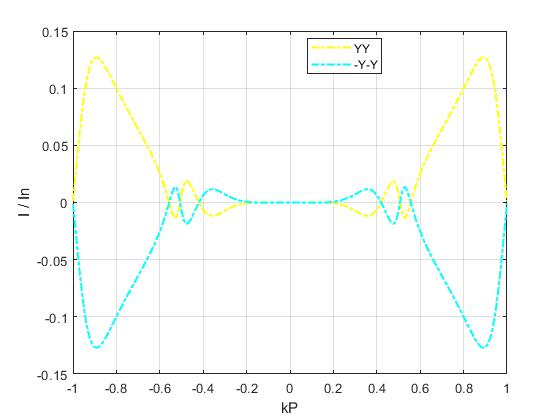}
	\includegraphics[scale=0.4]{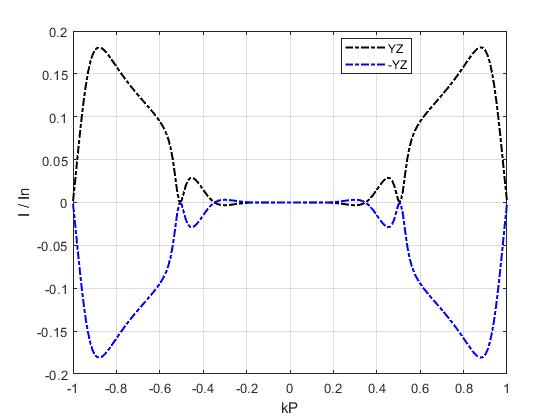}
	
	Figure 5.10: ZPC-$k_{P}$ for various geometries  \\($k_{F}d$=12,$\lambda$=0.6, Z=0.5)
\end{center}

\bigskip

\section{The effect of the magnetizations}

\par In the previous sections we studied the ZPC's dependence of our problem's various parameters, focusing on the basic magnetization geometries. Now, we shall examine the behaviour of the ZPC to the magnetizations' directions.
\par At first, we will study the ZPC that occurs, when both magnetizations are on one of the three basic planes, x-y and y-z planes, and "rotate" separately one from the other (the x-z plane does not include a y component and thus, the ZPC in that case will be zero). So, we will have a 3-D plot in which the independent variables will be the two angles (polar or azimuthal, according to the plane), while the dependent one will be the ZPC.
\par So, for the x-y plane we have: $\theta_{L}$=$\theta_{R}$=$90^o$ \& $\phi_{L}$,$\phi_{R}$$\in[0,360^o]$, where $\theta_{\nu}$ and $\phi_{\nu}$ are the polar and the azimuthal angle of each magnetization ($\nu=L,R$).
\par For the y-z plane: ($\phi_{L}$=$\phi_{R}$=$90^o$ or $\phi_{L}$=$\phi_{R}$=$270^o$)  \& $\theta_{L}$,$\theta_{R}$$\in[0,180^o]$.
\bigskip
\par In the figures following we represent the corresponding graphs:

\begin{center} 
	\includegraphics[scale=0.4]{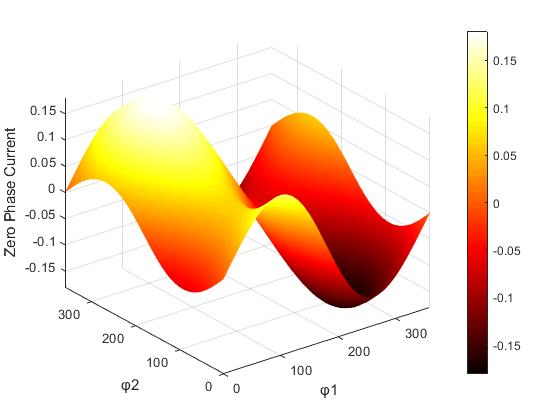}\includegraphics[scale=0.4]{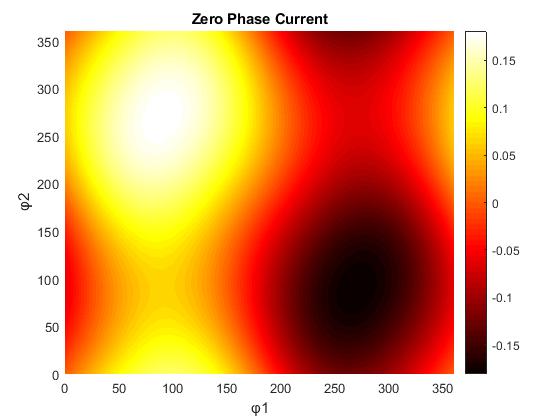}
	
	Figure 5.11: 3-D plot of ZPC as a function of $\phi_{L}$ \& $\phi_{R}$, x-y plane \\($\lambda$=0.6, Z=0.5, $k_{F}d$=12)
\end{center}
\par In this figure, we can see that the maximum appears for the Y-Y geometry and the minimum for the -YY. We can also observe that the inversion of the two angles changes the sign of the ZPC: $\phi_{L}\leftrightarrow\phi_{R}\Rightarrow ZPC\rightarrow-ZPC$.
\par Next, we have the graphs for the YZ plane. The left column is for $\phi_{\nu}=90^o$ and the right for $\phi_{\nu}=270^o$:
\newpage

\begin{center} 
	\includegraphics[scale=0.4]{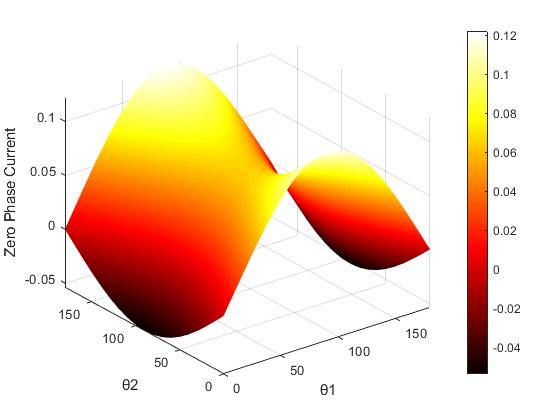}\includegraphics[scale=0.4]{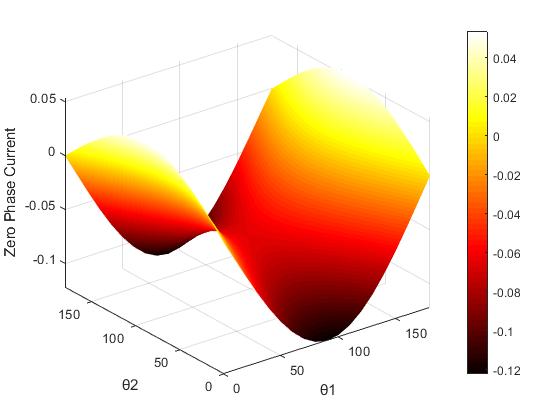}
	\includegraphics[scale=0.4]{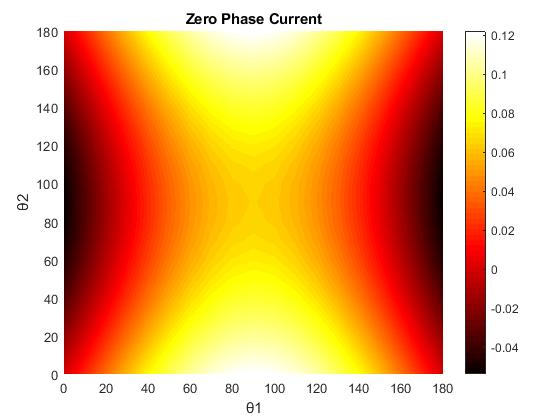}\includegraphics[scale=0.4]{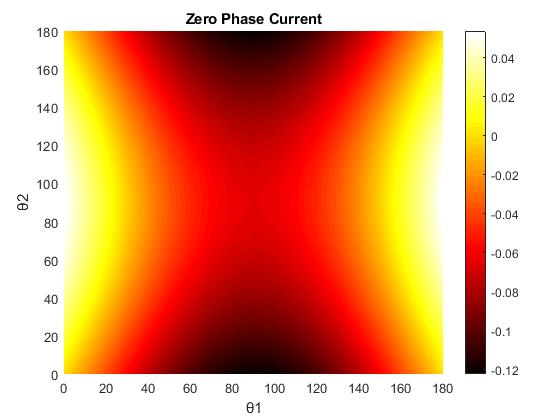}
	
	Figure 5.12: 3-D plot of ZPC as a function of $\theta_{L}$ \& $\theta_{R}$, y-z plane \\($\lambda$=0.6, Z=0.5, $k_{F}d$=12)
\end{center}
\par The first thing to observe in this figure, is the fact that the two graphs are antisymmetric: $ZPC_{1}(\theta_{1},\theta_{2})=-ZPC_{2}(\theta_{1},\theta_{2})$. In addition, the ZPC remains unchanged under the transformation of each (or both) of the angles: $\theta_{i}\rightarrow\pi-\theta_{i} (i=1,2)$. The maximum (minimum) for the first (second) set of graphs, appears for the Y-Z \& YZ geometries and the minimum (maximum) for ZY \& -ZY. 
\chapter*{Conclusions}
\addcontentsline{toc}{chapter}{Conclusions}
In the epilogue of this thesis we will summarize the most significant results of our work and discuss further possible expansions of this work. So first, we solved the Bogoliubov-de Gennes equations for an S/2DEG/S junction with two thin insulating and spin active interfaces, in the clean, ballistic limit. Next, we studied the current-phase relation of the junction for various values of the system's parameters (temperature, effective mass, spin-orbit coupling constant, junction's length, normal scattering strength, spin-flip scattering strength and direction). We also studied the dependence of the critical current of the junction from these parameters and examined the conditions under which it maximizes, as well as the contribution to this value for the different directions of the incident particles' wavevectors. Finally, we observed which conditions allow the appearance of the zero phase difference supercurrent of the junction, as well as how it varies to the change of our parameters.
\par More specifically, in chapter 3, we saw that when the spin-flip scattering interfaces' strength is set 0, the current phase relation is sine-like and decreases uniformly if we raise the values of the temperature, the length of the junction, the spin-orbit constant or the normal scattering strength, as well as if we decrease the effective mass. When the interfaces become spin active, the current-phase relation is no longer sinusoidal and a second harmonic appears. The amplitude of the second harmonic oscillates to the change of the spin-orbit constant and the junction's length. In addition, if we change the normal scattering amplitude, we can achieve a tunable 0-$\pi$ transition, as well. We also found that specific changes in the interfaces' magnetization geometries lead to the appearance of symmetries in the respective current-phase relations.
\par Following, in chapter 4 we found that the critical current of the junction rapidly vanishes as we raise the temperature or the normal scattering strength, as expected, while it decays in a slower rate to the increase of the junction's length or the spin-orbit constant, also showing oscillations which become sharpened for larger normal scattering strengths and spin-orbit constants. We also observed that, in general, the particles with almost incident wavevector directions are the ones that contribute mostly to the critical current. In addition, we saw that, for spin-active interfaces, the critical current depends not only from the difference, but from the sum of the magnetization angles, as well, with complicated expressions, due to the presence of the spin-orbit coupling.
\par Finally, in chapter 5, we found that, in order for the zero phase difference supercurrent to be non-zero, we must apply at least one of the two interface magnetizations on the y axis. The ZPC becomes maximized for specific values of the scattering strength (both normal \& spin-flip), the junction's length and the spin-orbit constant, while it also oscillates to the increase of these last two parameters. Another important notice, is that, contrary to the case of the critical current, only the particles with wavevectors almost parallel to the interface contribute to the ZPC, while the contribution of the incident wavevectors is zero in all the cases. We also studied the dependence of the ZPC from the interfaces' magnetization directions and found the general behavior as well as a number of symmetry relations.
\par There is a number of issues that we need to expand further in order to get a clearer understanding of the topic. First of all, we need to understand the theoretical background of the symmetry relations we discussed in chapter 3, which can be accomplished using the scattering amplitudes method. Another issue that needs further explanation is the oscillating behavior of the supercurrent as a function of the junction's length and the spin-orbit constant, as well as the 0-$\pi$ transition that we showed in chapter 3. In addition, we could also explain the current's dependence from the magnetizations' angles, seen in chapter 4, using again the scattering amplitudes method. Finally, we need to explain the appearance of the ZPC, seen in chapter 5, as well as it's dependence from the various parameters that we studied in the same chapter.

\medskip 

	\bibliographystyle{apsrev}
	\bibliography{sample}
	\addcontentsline{toc}{chapter}{Bibliography}

\end{document}